\pdfoutput=1
\documentclass[12pt,a4paper]{article}
\usepackage{a4wide}

\usepackage{subcaption}
\usepackage[english]{babel}
\usepackage{amsmath}   
\usepackage{graphicx}  
\usepackage{bm}        
\usepackage{amssymb}   
\usepackage{epsfig}
\usepackage{epstopdf}
\usepackage{amsmath}
\usepackage{fontenc}
\usepackage[toc]{appendix}
\usepackage{color,soul} 
\usepackage{datetime}
\usepackage{physics}
\usepackage{enumerate}
\usepackage{mdframed} 

\usepackage{footmisc}
\usepackage{textpos}
\usepackage{caption}
\usepackage{boxhandler}

\usepackage{xparse}
\usepackage{xifthen}
\usepackage{xstring}
\usepackage{xcolor}

\usepackage{tensor}
\newcommand{\ind}[1]{\indices{#1}}

\usepackage{tikz}

\usepackage{hyperref}

\newcommand{\be}{\begin{equation}}
\newcommand{\ee}{\end{equation}}
\newcommand{\bea}{\begin{eqnarray}}
\newcommand{\eea}{\end{eqnarray}}
\newcommand{\bean}{\begin{eqnarray*}}
\newcommand{\eean}{\end{eqnarray*}}

\newcounter{exercise}[subsection]
\renewcommand{\theexercise}{\arabic{exercise}}

\NewDocumentEnvironment{exercise}{s O{}}{ \refstepcounter{exercise} \noindent \textbf{Exercise \theexercise}\IfBooleanT{#1}{}\textbf{:}
\ifx&#2&\else\textbf{#2}.\fi
    \itshape}{}

\newenvironment{exparts}
   {
   
   \begin{enumerate} \setlength{\leftskip}{1em}}
   {\end{enumerate}\setlength{\leftskip}{0em}}
   
   \newcounter{exsol}[subsection]
\renewcommand{\theexsol}{\arabic{exsol}}

\NewDocumentEnvironment{exsol}{s O{}}{ \refstepcounter{exsol} \noindent \textbf{Solution to Exercise \theexsol}\IfBooleanT{#1}{}\textbf{:}
\ifx&#2&\else\textbf{#2}.\fi
    \itshape}{}

\newenvironment{solparts}
   {
   
   \begin{enumerate} \setlength{\leftskip}{1em}}
   {\end{enumerate}\setlength{\leftskip}{0em}}

\begin{document}




\begin{center}
\begin{center}
{\Large \bf{Modave Lectures on  Horizon-Size Microstructure,  Fuzzballs and Observations}} \medskip \\
\end{center}
\vspace{2mm}

\centerline{{\bf Daniel R. Mayerson}}
\vspace{5mm}

\centerline{Institute for Theoretical Physics,}
\centerline{KU Leuven, Celestijnenlaan 200D}
\centerline{B-3001 Leuven, Belgium}
\vspace{1mm}

%
{\footnotesize\upshape\ttfamily daniel.mayerson @ kuleuven.be} \\

\vspace{3mm}\bigskip
 
\textsc{Abstract}

\end{center}
 
\vspace{-3mm}
\noindent These lecture notes discuss various aspects of the fuzzball paradigm, microstate geometries, and their role in gravitational phenomenology. 
We briefly introduce the information paradox and discuss how the fuzzball paradigm aims to resolve it. Then, some important families of fuzzball solutions in supergravity, called microstate geometries, are explored: multi-centered bubbling geometries and superstrata. Finally, we will review some very recent developments of the phenomenology of fuzzballs and delineate the exciting opportunities as well as the limitations of studying fuzzballs as alternatives to black holes in current and future observations.








\subsection*{A guide to these lecture notes}
These notes are a brief introduction to fuzzballs, microstate geometries, and their role as compact objects in gravitational phenomenology for precision black hole observations. They are based on lectures I gave at the XVII Modave Summer School in Mathematical Physics in September 2021.

These notes are emphatically \emph{not} meant as an alternative to more comprehensive lecture notes, such as \cite{Warner:2019jll} and \cite{Bena:2007kg} for multi-centered bubbled geometries, or \cite{Shigemori:2020yuo} for superstrata. Rather, if \cite{Warner:2019jll,Bena:2007kg,Shigemori:2020yuo} are the ``manual'' for these geometries, then these notes should be seen as the ``Quick start guide'': a practical collection of some of the most pertinent material that one needs to understand and \emph{start working} with these geometries. (Note that, despite the length of this entire document, the main part --- Section \ref{sec:FBparadigm} introducing fuzzballs and Section \ref{sec:multicenter} discussing multi-centered geometries --- is under 20 pages.)

It is also not necessary to go through these entire notes, or even to go through each section sequentially; the reader can pick and choose the topics which they are interested in learning about.

These notes are also complementary to my review \cite{Mayerson:2020tpn} on ``Fuzzballs \& Observations'', which is an overview of fuzzball phenomenology, meant to introduce the relevant concepts and ideas (both in fuzzballs and in phenomenology) without too many technical details of the geometries. By contrast, these notes give precisely the minimal technical details necessary to actually start performing concrete calculations with microstate geometries.

\textbf{Section \ref{sec:FBparadigm}} introduces the fuzzball paradigm, discussing mechanisms and concepts that lie at the basis of the existence of microstructure. The multi-centered bubbling geometries are derived and discussed in \textbf{Section \ref{sec:multicenter}}. Superstrata are discussed in \textbf{Section \ref{sec:SS}}, albeit with quite a bit less detail. \textbf{Section \ref{sec:obs}} discusses applying  fuzzballs and horizon-scale microstructure to observations and gravitational phenomenology. 

The appendices collect some additional information. \textbf{Appendix \ref{sec:stringtheory}} is a brief overview of some concepts in string theory that arise frequently in the discussion of fuzzballs. \textbf{Appendix \ref{sec:summarybubble}} is a reference containing all the necessary information to construct multi-centered solutions (from Section \ref{sec:multicenter}). \textbf{Appendix \ref{sec:exercises}} is a collection of six exercises (including solutions), most on multi-centered geometries and some of their more important properties.

\setcounter{tocdepth}{2}
\vspace*{-1\baselineskip}
\tableofcontents

\newpage

\section{The Fuzzball Paradigm}\label{sec:FBparadigm}

In this Section, we will describe the fuzzball paradigm, its motivation, and some of its aspects and limitations. The main relevant references (and suggested further reading) are: \cite{Mathur:2009hf} for Section \ref{sec:infoparadox}; \cite{Strominger:1996sh,Mathur:2005zp} for Section \ref{sec:SV}; \cite{Bena:2007kg} (especially Section 6 therein) for Section \ref{sec:microstructure}; \cite{Kraus:2015zda} for Section \ref{sec:formation}; and \cite{Mayerson:2020tpn} (especially Section 2.3 therein) for Section \ref{sec:limitations}.

\subsection{Introduction: The information paradox}\label{sec:infoparadox}

A black hole can be formed by the violent collapse of matter, or the merging of dense objects. A horizon forms and grows, masking the collapsing or merging matter behind it. After its formation, the resulting black hole will then quickly relax to a stationary state. Various uniqueness theorems in general relativity tell us that the resulting stationary black hole is only characterized by its mass and angular momentum (and possibly charge).

\begin{figure}[ht]\centering
 \includegraphics[width=0.52\textwidth]{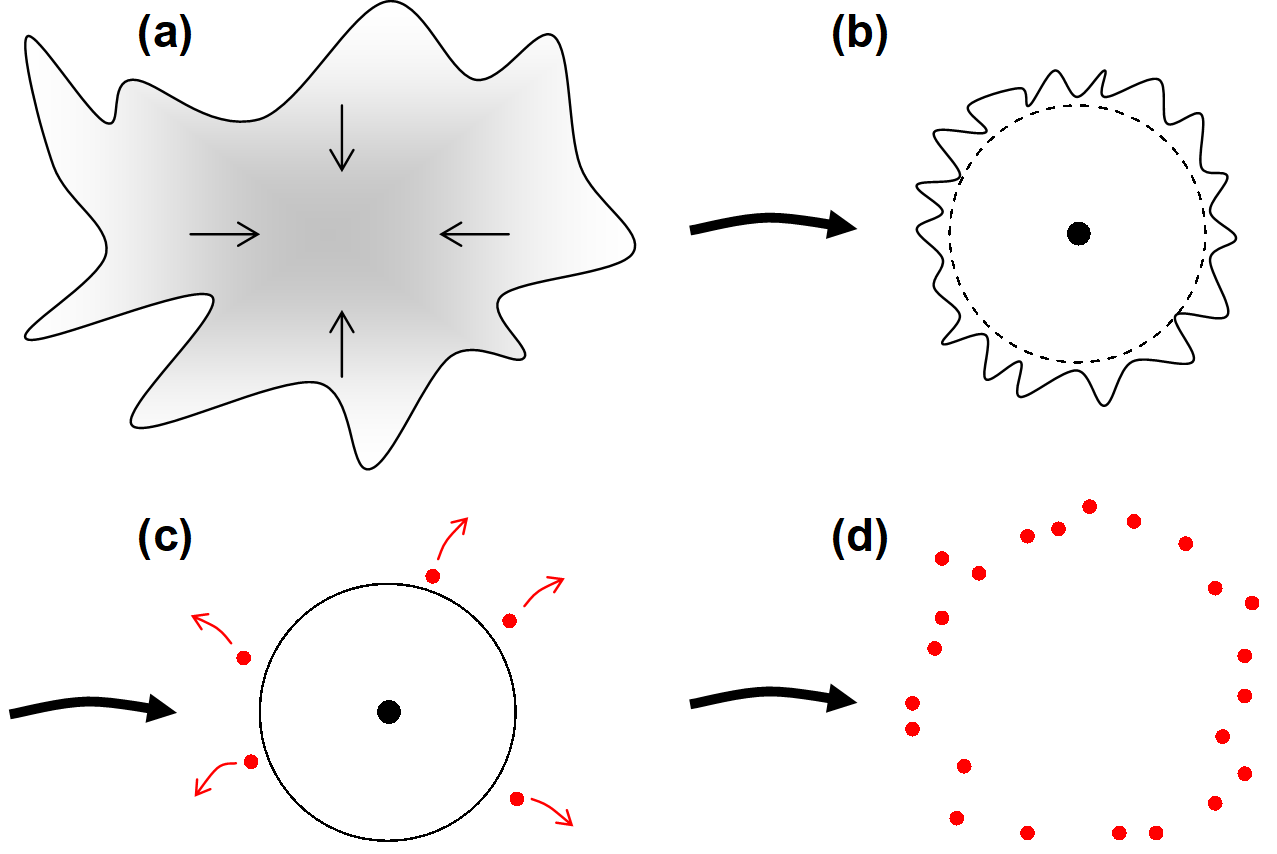}
 \caption{The naive black hole evolution picture which leads to the information paradox. (a) Matter collapses. (b) A horizon is formed (the dotted line) and the new black hole relaxes to a stationary state; the singularity at the center has also already formed. (c) Once in its (classically) stationary state, the black hole only changes by radiating thermal Hawking particles (in red) from its horizon region. (d) After a long time, the black hole has entirely evaporated, and we are left with only the informationless Hawking radiation.}
 \label{fig:BHevolution}
\end{figure}

If the universe was classical, the story of the black hole's evolution would end here. However, quantum mechanically, the black hole has a temperature and emits \emph{Hawking radiation}. We can derive the presence and properties of this radiation by considering quantum fields on a (fixed!) black hole background. Since the radiation is thermal, it is entirely featureless and does not contain any information: its temperature is solely determined by the properties of the black hole --- from black hole uniqueness, this means just its mass and angular momentum. Hawking radiation is a very small effect, but if we imagine waiting for a very long time, the black hole should fully evaporate into this thermal, informationless radiation. We are left with no black hole, and only thermal radiation that contains no information. See Fig. \ref{fig:BHevolution} for a depiction of the black hole evolution.

\emph{Where did the information from inside the black hole go?} In quantum mechanics, time evolution should be unitary: if we know the final state, we should (at least in principle) be able to reconstruct the initial state. Here, the final state is just thermal radiation; it does not distinguish between the many different initial states from which the black hole could have formed. Hawking radiation seems to suggest \emph{information loss}, whereas unitarity in quantum mechanics preserves information. This is the \emph{information paradox} for black hole horizons.

 Clearly, something must be \emph{wrong} with the above picture in order for unitarity to be preserved and information loss to be prevented. A first logical guess would be that Hawking's calculation can receive small corrections in a full quantum gravity calculation, where also the black hole background geometry changes as it radiates. Unfortunately, performing such a calculation is not quite in our reach. However, we do have Mathur's theorem, which tells us that \emph{small corrections to Hawking radiation are not enough} to prevent information loss. No matter what the full quantum gravity calculation would be, it must either result in information loss, or show that there are \emph{large} corrections to physics at the horizon scale of the black hole.\footnote{Note that Hawking radiation is all about the horizon-scale physics. The ``normal'' horizon-scale physics can be summarized by the assumption that the horizon in general relativity is ``not a special place''; it is in a vacuum state from the point of view of an infalling observer. This is essentially the main ingredient needed to be able to derive that Hawking radiation is being emitted from the black hole.}

\begin{figure}[ht]\centering
 \includegraphics[width=0.5\textwidth]{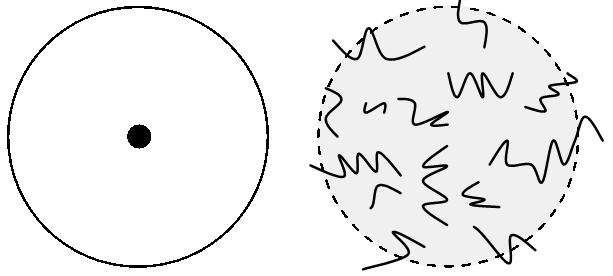}
 \caption{A black hole (left), with a horizon and a singularity at its center. A fuzzball (right), by contrast, has no singularity and no horizon, but instead consists of quantum, stringy ``fuzz'' that extends all the way to the would-be horizon scale.}
 \label{fig:BHfuzzball}
\end{figure}

Specifically, following Mathur \cite{Mathur:2009hf,Mathur:2017fnw}, there are roughly three options that are possible to resolve the information paradox:
\begin{itemize}
 \item \emph{The existence of remnants:} After a long time of radiating, the black hole becomes Planck sized.  At this point, it is too small to be reliably described by a geometry in general relativity --- quantum fluctuations on this geometry would be as large as the object itself. Perhaps Hawking's radiation breaks down entirely there, and the left-over ``remnant'' black hole stops radiating. This remnant could then carry all the information that was originally contained in the black hole. In this picture, the resolution to the information paradox is that information is trapped in the remnant, but not lost.
 
 While logically possible, this remnant scenario is not very attractive: it would require that a Planck sized object is able to carry an arbitrarily large amount of information (since the initial black hole was arbitrarily large). There are also arguments from AdS/CFT that indicate this scenario is unlikely or even impossible \cite{Mathur:2017fnw}.
 \item \emph{Non-local corrections:} Mathur's theorem tells us that small, \emph{local} corrections to the physics at the horizon scale are not enough. This does not preclude the possibility of non-local corrections in quantum gravity --- however small --- that could alter the near-horizon physics in such a way that the information paradox is resolved.
 
 String theory, and especially holography, has given us plenty of evidence that quantum gravity must have inherent non-localities \cite{Harlow:2018fse}. These often manifest in subtle ways, so that a semi-classical (local) limit is still possible without contradictions. 
 There are also a number of proposals for how non-local quantum corrections at the horizons of black holes could manifest themselves. For example, the Papadodimas-Raju paradigm \cite{Papadodimas:2012aq,Papadodimas:2013jku,Papadodimas:2013wnh,Papadodimas:2015jra} is a concrete calculation in holography for a large, asymptotically AdS black hole. Here, it can be shown in a rigorous fashion that the degrees of freedom behind the horizon are essentially a non-local reshuffling of the degrees of freedom outside the horizon. More recently, the ``island'' proposal \cite{Almheiri:2019yqk,Penington:2019npb,Almheiri:2019hni,Almheiri:2019psf,Bousso:2022ntt} is based on similar ideas. There is also Giddings' proposal of ``non-violent non-locality'' \cite{Giddings:2006sj,Giddings:2012bm,Giddings:2013kcj,Giddings:2014ova,Giddings:2016plq,Giddings:2019ujs}.
 \item \emph{Large corrections at the horizon scale:} If \emph{small} corrections to horizon-scale physics are not enough to resolve the information paradox, then there must be \emph{large} corrections!
 \end{itemize}
  The latter option is the heart of the fuzzball proposal: we replace the horizon by \emph{something else entirely} --- specifically, by quantum, stringy fuzz that exists and manifests itself at the horizon scale (see Fig. \ref{fig:BHfuzzball}). If we look at the fuzzball from far away, we should see something that looks very much like a black hole; only when we get close enough to the would-be horizon to probe the actual fuzzy structure that exists there, will we start seeing measurable differences from black hole physics.\footnote{However, note that the idea of ``fuzzball complementarity'' would imply that only high-energy observers would be able to ``see'' the effects of the microstructure and have a different experience from the usual black hole infall \cite{Mathur:2012jk,Mathur:2010kx,Mathur:2011wg,Mathur:2012zp,Mathur:2012dxa}.}
  
  It is useful to point out that the above options of resolving the information paradox ---  non-locality and large corrections such as fuzzballs\footnote{Firewalls \cite{Almheiri:2012rt} are also an example of introducing large corrections at horizon scales.} --- are usually phrased as exclusive options, where either one or the other can be correct, but not both. My opinion is that the truth may be more subtle; it is quite possible that the different ideas are not completely mutually exclusive, but rather highlight or approach different aspects of the same issue. For example, the complete lack of horizon in fuzzballs could be seen as interpreting the ``interior'' degrees of freedom of the black hole as a non-local ``reorganizing'' of the degrees of freedom on the ``exterior'' --- this starts to sound like non-locality. In any case, no matter which paradigm has your preference, it is certainly always meaningful to find out what can be learned, reinterpreted, or understood from alternative approaches to resolving the information paradox.

  \subsection{Microstates and ensembles}\label{sec:SV}
  Let's put our musings on the information paradox aside for a moment, together with our expected large corrections at the black hole horizon scale. Instead, let's turn our focus to black hole \emph{entropy}. It is well-known that black holes must have an entropy $S_\text{BH}$, and moreover that this entropy must be proportional to its area:\footnote{If a black hole horizon did not carry entropy, then the irreversibility of entropy increase would be violated when we throw something with a finite entropy into a black hole. See the introduction of \cite{Mathur:2005zp} for more discussion; the original argument is from Bekenstein \cite{Bekenstein:1973ur}.}
  \be S_\text{BH} = \frac{A_\text{BH}}{4G_N}.\ee
This is a very curious formula. In conventional systems, entropy is an extensive quantity that scales with the \emph{volume} of space instead of the \emph{area} (of the space's boundary). The situation becomes even more curious when we consider the microscopical implications. In the microcanonical ensemble at a given energy, statistical mechanics tells us that thermodynamic entropy $S_\text{th}$ arises from counting the number $N_\text{micro}$ of different microscopic states that are in this ensemble: $S_\text{th} = k_B \ln N_\text{micro}$. So if black holes have entropy, then we should be able to interpret a black hole as a thermodynamic ensemble of microstates. \emph{What are these black hole microstates?}

  \begin{figure}[ht]\centering
  \hspace{2em}\begin{subfigure}[t]{0.45\textwidth}\centering
    \includegraphics[width=0.8\textwidth]{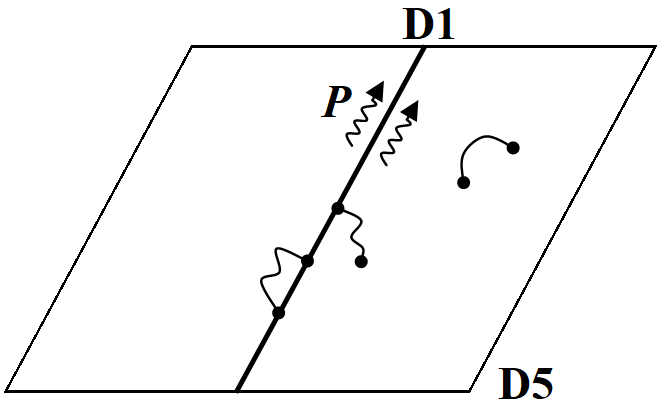}
    \caption{$g_s\ll 1$}
  \end{subfigure}\hfill
  \begin{subfigure}[t]{0.45\textwidth}\centering
    \includegraphics[width=0.5\textwidth]{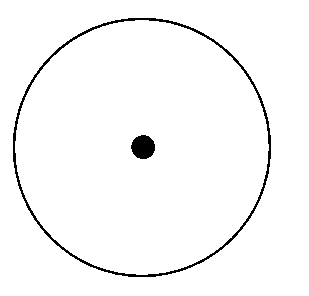}
    \caption{$g_s\gg 1$}
  \end{subfigure}\hspace{2em}
 \caption{The two dual pictures describing the D1-D5-P system at different couplings. At weak coupling $g_s\ll1$, the branes are rigid and the system is captured by an ensemble of CFT states which describes the string excitations on these branes. (The string endpoints can be both on the D1 brane, both on the D5 brane, or one on each the D1 and D5 branes.) At strong coupling $g_s\gg 1$, the system is described by the D1-D5-P black hole geometry in supergravity.}
 \label{fig:D1D5duality}
\end{figure}

Holography gives us some insight into this problem, with the famous Strominger-Vafa calculation \cite{Strominger:1996sh} --- this actually preceded Maldacena's seminal ``discovery'' of AdS/CFT holography \cite{Maldacena:1997re}, but in hindsight is a prime example of the holographic duality. Consider a stack of coincident D1 and D5 branes, which share a common direction along which we also allow a momentum charge P to run; see Fig. \ref{fig:D1D5duality}. We have $N_1, N_5, N_P$ units of quantized D1, D5, and P charge, respectively. We can consider this system either at low string coupling, $g_s\ll 1$, or strong coupling, $g_s\gg 1$; in either case, we have a description which captures its physics:
\begin{itemize}
 \item Strong coupling $g_s\gg 1$: The system can be described in supergravity by the D1-D5-P black hole. This is a black hole with entropy $S_\text{D1-D5-P} = 2\pi \sqrt{N_1 N_5 N_P}$. 
 \item Low coupling $g_s\ll 1$: A description of this system is obtained by considering the possible excitations of the strings that stretch between the different D1 and D5 branes. This gives rise to a effective description by a \emph{conformal field theory} called the D1-D5 CFT. This is a quantum field theory without gravity. In particular, we are interested in the states in this CFT that correspond to a given momentum charge $N_P$. Counting these states, we find that there are $\sim \exp \left(S_\text{D1-D5-P}\right)$ such states.
\end{itemize}
The holographic ``duality'' is then the realization that these two descriptions are really just two faces of the same coin --- one valid at weak coupling, and one at strong coupling.\footnote{To be able to relate the counting in the CFT at low coupling to the (black hole entropy) counting in supergravity at strong coupling, another crucial ingredient is the supersymmetry of this system. This ensures that the states that we are counting are protected as we move from low to high coupling \cite{Baggio:2012rr}.}

The Strominger-Vafa counting gives us some insight into what a black hole microstate should be. The D1-D5-P black hole in supergravity corresponds (using holography) to a thermal ensemble in the dual CFT. The dual of the black hole is then the \emph{whole ensemble} of microstates in the CFT. However, using the holographic correspondence in the opposite direction, we should also be able to identify a gravitational dual for each \emph{individual microstate} in the CFT. \emph{What are these (gravitational) black hole microstates?}

The answer is simple: each individual gravitational black hole microstate is a \emph{fuzzball}. (For simplicity, we can take this as our definition of ``fuzzball'' \cite{Bena:2013dka}.) Every one of these fuzzballs should be entirely \emph{horizonless}: if a fuzzball itself had a horizon, then it would again have a non-zero entropy (just like the black hole does) and so would not correspond to a single microstate. Once again, we are led to the conclusion that fuzzballs must have \emph{large} deviations at the (would-be) horizon scale of a black hole --- to ensure that fuzzballs are altogether \emph{horizonless}!

A black hole exists as a well-behaved object in (super)gravity. However, a priori there is no reason to expect or hope that all fuzzballs would similarly exist as geometric objects. In general, fuzzballs should and will be quantum, stringy objects that are not well described by a metric or anything resembling a (semi-)classical geometry. However, \emph{some} fuzzballs are semi-classical coherent states --- very much like the coherent states of the harmonic oscillator --- that are captured by geometric solutions in supergravity. These geometric fuzzballs must be horizonless and entirely smooth (i.e. without curvature singularities); such coherent fuzzballs are called \emph{microstate geometries} \cite{Bena:2013dka}. These geometries represent an exciting window into the microscopics of black holes, since we are able to write down these geometries explicitly in supergravity and study their properties.

\subsection{How is horizon-scale microstructure possible?}\label{sec:microstructure}
We have now formed an idea of black holes in string theory as thermodynamic ensembles of horizonless microstates or fuzzballs. As mentioned above, we certainly don't expect all such fuzzballs to be nice,  geometric objects --- but (hopefully) there should be some that can be described as a classical solution in some supergravity theory, i.e. microstate geometries. These geometries should look like a black hole from far away, but should not have a horizon and should instead exhibit some kind of ``microstructure'' when we get close to the (would-be) horizon scale.

Any object with mass or energy has a tendency to gravitationally contract. Buchdahl's theorem \cite{PhysRev.116.1027} tells us that well-behaved matter, under reasonable assumptions, has to extend to $(9/8) r_S$, where $r_S$ is its Schwarzschild radius --- if we compress the matter any further, it will unavoidably collapse and form a black hole. Microstate geometries should be much more compact, and have microstructure much closer to the would-be horizon scale. So how is such horizon-scale microstructure possible, without it wanting to collapse?\footnote{This is actually a problem for many exotic compact objects or horizonless black hole mimickers which do not arise from a top-down construction; they often suffer a trade-off between compactness and stability. See further in Section \ref{sec:obstofuzzpheno}.}

String theory provides remarkable mechanisms that support such horizon-scale microstructure. We will consider two closely related phenomena to give us some insight into what fuzzball microstructure is made  of.

\subsubsection{Branes dissolved in flux} \label{sec:branesflux}
What is a source?

This question likely evokes an image of a particle or other object that carries a given charge --- for example, an electromagnetically charged particle such as an electron. In this case, the object itself is a singularity (usually point-like, in the case of a particle) in the theory; a potential or field will diverge ($V\sim q/r$ or $E\sim q/r^2$) at the location of the charge. This divergence or singularity at the source tells us our theory breaks down at the location of the source and cannot be trusted arbitrarily close to it. In gravity, the analogue of such source singularities are curvature singularities such as those at the center of a black hole; these are natural places to expect large quantum and stringy corrections.

  \begin{figure}[ht]\centering
\includegraphics[width=0.8\textwidth]{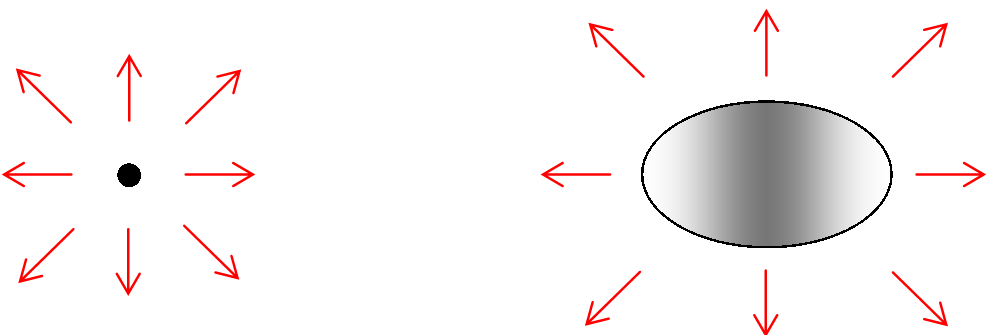}
 \caption{A localized, singular point source (left), and a ``source dissolved in flux'' (right).}
 \label{fig:sources}
\end{figure}

Branes in string theory are extended objects that carry certain string theoretic charges. For example, a D$p$-brane is an electric source for a Ramond-Ramond (RR) $(p+1)$-form potential and thus a $(p+2)$-form field strength. Solutions in supergravity that represent simple branes exhibit singularities both in these RR fluxes and in the spacetime curvature at the location of the branes.

However, branes can also undergo a remarkable phenomenon called a ``geometric transition'' \cite{Bena:2007kg}, where they become entirely ``dissolved in flux''; see Fig. \ref{fig:sources}. In essence, string theory can ``resolve'' brane singularities by changing the \emph{topology} of spacetime to carry non-trivial topological cycles with electromagnetic fluxes on them. This remarkable phenomenon ensures that the spacetime still carries the same brane charge (as can be calculated, say, by a Gaussian flux integral at infinity), but without this charge actually being localized anywhere. The charge is ``dissolved'' into the very geometry of the spacetime itself --- and as a result, the geometry is entirely smooth and without singularities.

\subsubsection{The supertube transition}\label{sec:supertube}
A closely related phenomenon is the \emph{supertube} transition. Consider a string (an ``F1-brane'') which has D0-brane ``beads'' located on it. (D0-branes are like ``particles'' with no spatial extent.) This is a perfectly acceptable, supersymmetric configuration in string theory, and is singular at the location of the string.

  \begin{figure}[ht]\centering
\includegraphics[width=0.7\textwidth]{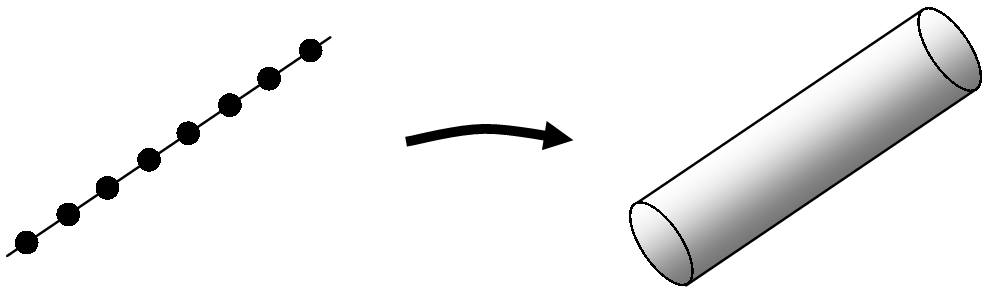}
 \caption{The supertube transition: on the left, we have a (singular) configuration of an F1-string with D0-brane ``beads'' along it. After the ``puff up'', we no longer have any localized F1 or D0 sources. Instead, we have a D2-brane dipole (here, with a circular profile) with electromagnetic fluxes on its worldvolume. These fluxes ensure the supertube carries the same F1 and D0 charges as in the left picture.}
 \label{fig:supertubetransition}
\end{figure}

However, another configuration that is allowed and carries the same (F1 and D0) charges is the \emph{supertube} \cite{Mateos:2001qs}: a D2-brane which is shaped like a cylinder and carries electromagnetic fluxes living on its worldvolume; see Fig. \ref{fig:supertubetransition}. Because the D2 brane is cylindrically shaped, there is actually no net D2-brane charge at infinity --- this configuration only carries D2 ``dipole'' charge. The electromagnetic fluxes on the D2 brane ensure that it does carry  F1 and D0 charges, but also has the added benefit of \emph{preventing the D2-brane from contracting}. If there would be no fluxes, the D2 brane would simply contract along the circular direction due to its own gravitational weight. However, the fluxes on the brane's worldvolume keep it from doing so --- the electromagnetic repulsion precisely counteracts the gravitational attraction. Said differently, the electromagnetic fields on the D2-brane worldvolume give the D2 brane an intrinsic angular momentum (of the electromagnetic $\vec{E}\times \vec{B}$ type), and the resulting centrifugal force of the spinning D2 brane counteracts the gravitational attraction.

This transition from branes to a supertube carrying brane charge is sometimes colloquially called a ``puff up''. 
Note that in the puffed up D2 supertube, the F1- and D0-brane charges are no longer localized: they are entirely ``dissolved'' in the electromagnetic fluxes on the D2 brane. The supertube transition is thus precisely an example of branes dissolving in flux in string theory! We will revisit this supertube puff up transition in a different brane system in Section \ref{sec:stringmomentum}.

\subsubsection{Horizon-scale microstructure}
The supertube transition described above for the F1-D0 system is applicable for many other combinations of branes. In general, when we put multiple kinds of branes together, they tend to want to ``puff up'' and dissolve themselves into some other object (like the supertube), which can itself also dissolve into geometry.

Such puff-ups and delocalized brane sources dissolved in flux are at the heart of the horizon-scale microstructure in smooth microstate geometries. These geometries have no singularities as they only carry branes that are dissolved in fluxes; the geometries are completely smooth. They do carry (brane) charges, but these are not localized. The brane and charges have ``puffed up'' to \emph{horizon-sized} topological ``bubbles''. This bubble structure is entirely stable from collapse and can be made arbitrarily compact (i.e. can sit as close to the would-be horizon as we like); the electromagnetic fluxes on the bubbles keep them from collapsing. Voil\`a:  horizon-scale microstructure!

We will discuss such topological bubble structures further in Section \ref{sec:horizonscalebubbles} in the context of the multi-centered bubbling microstate geometries.

\subsection{How do we form fuzzballs (and not black holes)?}\label{sec:formation}

Picture a large object undergoing gravitational collapse. In the ``standard'', classical picture, unless stopped by other repulsive forces (such as fusion pressure inside a star), the object will continue on collapsing and eventually form a horizon, creating a black hole. When this horizon initially forms, there are no large curvatures yet anywhere --- in particular, there is not yet a singularity at the center of the object --- and so the usual, effective field theory viewpoint tells us that quantum corrections to this picture can only be small. There does not seem to be any obstacle against forming a horizon.

By now, we understand that the fuzzball paradigm tells us that this effective field theory viewpoint may not be entirely valid. To resolve the information paradox, fuzzballs introduce counter-intuitively \emph{large} corrections at the horizon scale that make the actual fuzzball entirely horizonless.  But how can we avoid the \emph{formation} of horizons in the first place, for example in such a gravitational collapse scenario? This is a \emph{dynamical} question and requires new insights since we have so far only discussed fuzzball states without dynamics.

The key ingredient that prevents the actual formation of a horizon is that of a \emph{large phase space} of states which opens up when an object becomes horizon-sized \cite{Mathur:2008kg,Mathur:2009zs,Kraus:2015zda}.
Precisely when the collapsing matter has almost all ``entered'' the radius of its would-be horizon, a huge phase space of possible states becomes available --- the number of microstates of the corresponding black hole $\sim e^{S_\text{BH}}$. Because this number of states is \emph{so large}, quantum effects can become very large and invalidate the (effective field theory) classical picture of horizon formation. In particular, the collapsing object can \emph{quantum tunnel} into a fuzzball state with a probability that can be estimated as $\sim e^{-S_\text{BH}}$ \cite{Kraus:2015zda} --- an extremely small probability, which is why we typically do not care about quantum tunneling effects for large objects in classical physics. But because  there are $\sim e^{S_\text{BH}}$ such fuzzball microstates available to tunnel into, the \emph{total} probability to tunnel into \emph{any} fuzzball state is $\sim e^{-S_\text{BH}}\times e^{S_\text{BH}} \sim \mathcal{O}(1)$ --- so the collapsing object will \emph{definitely} quantum tunnel into fuzzballs, and will \emph{not} continue its classical collapse to form a horizon! 

It would be nice to calculate such tunneling rates in actual fuzzball states to confirm this argument quantitatively. Unfortunately, such general quantum dynamical calculations are well beyond the scope of the current methods, except in very controlled, special setups where some explicit calculations can be done (and seem to confirm this argument) \cite{Bena:2015dpt}.  However, the heuristic argument above does at least show the principle of how fuzzballs can be \emph{dynamically} formed \emph{instead of horizons}.

\subsection{Limitations of microstate geometries}\label{sec:limitations}
As mentioned above, microstate geometries are the special, semi-classical coherent fuzzball states that we can represent and study as geometric solutions in a supergravity theory. While immensely interesting objects, it is also important to understand  the inherent restrictions of such microstate geometries. I will touch on a few important limitations here; see also e.g. Section 2.3 of \cite{Mayerson:2020tpn} for further discussion and references on these and other limitations.

Finding solutions in supergravity theories is typically a difficult problem, involving solving multiple coupled non-linear, second order partial differential equations. Demanding that a solution is supersymmetric often makes life easier, as this only gives first-order equations to solve. As a consequence, most of the microstate geometries that have been constructed are \emph{supersymmetric}, such as the multi-centered bubbling geometries we will discuss below.

It is also sometimes possible to cleverly set up a system of first-order equations, reminiscent of supersymmetry, but which gives solutions that are not supersymmetric \cite{Bena:2009ev}. Such solutions are still ``extremal'', which means the corresponding black hole has zero temperature. This means the corresponding black hole has the maximal allowed charge(s) and/or angular momentum.

Realistic black holes are certainly not supersymmetric, and should also be non-extremal, and probably also completely uncharged. The supersymmetric or non-supersymmetric but extremal microstate geometries that we have available are then not very realistic as astrophysical objects. This is an important limitation of most microstate geometries. (Note that very recently, there has been some exciting progress in constructing certain classes of new non-extremal microstate geometries \cite{Bah:2020ogh,Ganchev:2021pgs}.)

Dynamics is also a difficult issue. How do fuzzballs or microstate geometries \emph{form}? How do they evolve? How do they behave under perturbations? Finding the (stationary) microstate geometry solution itself is often already a difficult problem; moving away from stationarity to study dynamics would add a whole new layer of complexity. Only a few such issues have been attacked so far. For example, the (analogue of) Hawking radiation in certain fuzzballs can be calculated \cite{Chowdhury:2007jx}, suggesting that fuzzballs mimic black hole Hawking radiation by a complicated process of emission from their inherent unstable modes.

Finally, an important caveat of microstate geometries is that of \emph{typicality}. Microstate geometries are solutions in supergravity and represent semi-classical coherent microstates of a corresponding black hole. These microstate geometries only make up a small fraction of the total phase space of possible microstates of the black hole. If we study microstate geometries and their properties, how well does this capture properties of generic, \emph{typical} microstates? It is possible that the \emph{atypicality} of microstate geometries means they will sometimes behave in ways that are very different from most (typical) microstates of the black hole \cite{Raju:2018xue}. Luckily, there are some arguments that suggest that microstate geometries will indeed capture certain features of typical states, such as energy gaps \cite{Bena:2007qc}.

It is important to keep these limitations of microstate geometries in mind when studying them. However, despite their downsides, my viewpoint is that microstate geometries still represent useful tools that can point the way towards understanding features of black hole microstates, and in particular horizon-scale microstructure that replaces horizons. Many insights have already emerged from constructing and analyzing microstate geometries, both theoretical (e.g. in precision holography) and more practical (e.g. in phenomenology), where microstate geometries are used as tools and models that point to interesting observables that can show the existence of horizon-scale microstructure in observations.

\section{Multi-centered Bubbling Geometries}\label{sec:multicenter}
We have waved our hands around enough; time to get them dirty with some real calculations. We will discuss the construction of multi-centered bubbling geometries in this section. We will construct them in five-dimensional supergravity, although we will show how they can ``live'' both in five- or four-dimensional asymptotics --- the latter being phenomenologically preferred, of course. These geometries are supersymmetric, and are microstates of the supersymmetric three-charge, rotating BMPV black hole in five dimensions, or of the supersymmetric eight-charge, static (non-rotating) black hole in four dimensions.
These microstate geometries are often called Bena-Warner geometries \cite{Bena:2004de,Berglund:2005vb,Bena:2007kg}, or Denef-Bates geometries \cite{Denef:2000nb,Denef:2002ru,Bates:2003vx} in a four-dimensional perspective.

The main reference for this entire section is \cite{Bena:2007kg}. See therein especially Section 3.2 (for the supersymmetry equations); Section 4.1 about Gibbons-Hawking metrics (for Section \ref{sec:GHmetrics} below); Section 5.1 for solving the supersymmetry equations (for Section \ref{sec:susyeqs} below); Section 5.3 about CTCs (for Section \ref{sec:bubbleeqs} below); Sections 6.2-4 for the smooth, bubbling solutions, the bubble equations, and five-dimensional asymptotic charges (for Sections \ref{sec:bubbleeqs} and \ref{sec:moduli} below); and Sections 1.1 and 2.1 about brane interpretations (for Section \ref{sec:braneinterpret} below).
Finally, see \cite{Gibbons:2013tqa} (and also Section 8.1 in \cite{Bena:2007kg}) for further discussion on the horizon-scale microstructure of topological bubbles of Section \ref{sec:horizonscalebubbles} below.

Another great reference for the construction of these geometries is the lecture notes \cite{Warner:2019jll}, especially Sections 3, 4 and 6.1 therein; additionally, Section 5 is relevant for scaling geometries as discussed in Exercise \ref{ex:scaling}.

\subsection{Setup: action and fields}
We will consider a five-dimensional supergravity theory that comes from a simple toroidal compactification of string theory, although it is possible to generalize to other compactifications. The fields in the theory are the metric $g_{\mu\nu}$, three gauge fields $A^I$ (with $I=1,2,3$) with corresponding field strengths $F^I = dA^I$, and three constrained scalar fields $X^I$. The three scalar fields satisfy a single algebraic constraint, so there are really only two scalar field degrees of freedom.

The bosonic field action is given by:
\be \label{eq:5Dlagr} S_\text{5D} = \int d^5x\, \left[\sqrt{-g}\left( R - \frac12 Q_{IJ} F\ind{^{I\, \mu\nu}} F\ind{^J_{\mu\nu}} - Q_{IJ} \partial^\mu X^I \partial_\mu X^J\right) - \frac{1}{24}C_{IJK}\epsilon^{\mu\nu\rho\sigma\lambda} A_\mu^I F_{\nu\rho}^{J} F_{\sigma\lambda}^{K}\right],\ee
with the definitions:
\begin{align}
 Q_{IJ} &\equiv \frac92 X_I X_J - \frac12 C_{IJK} X^K, & 
 X_I &\equiv \frac16 C_{IJK} X^J X^K.
\end{align}
The action contains kinetic terms for the scalars and gauge fields, where the latter one couples the scalars to the gauge fields; in addition, there is a Chern-Simons ``$A\wedge F\wedge F$'' term involving (only) the gauge fields.

The algebraic constraint that the scalars $X^I$ need to satisfy is:
\be  \frac16 C_{IJK} X^I X^J X^K = 1.\ee
The simple toroidal compactification we are considering means that we take:
\be C_{IJK} = |\epsilon_{IJK}|.\ee
This implies that $Q_{IJ} = (1/2)\delta_{IJ}(X^I)^{-2}$ (no sum over $I$). 
We can parametrize the scalars by three unconstrained functions $Z_I$ as:
\be \label{eq:5Dscalars} X^I = \frac{Z}{Z_I}, \qquad Z  \equiv (Z_1Z_2Z_3)^{1/3}.\ee

\subsection{Imposing supersymmetry: the linear system}\label{sec:susyeqs}
Instead of solving the equations of motion that come from the Lagrangian (\ref{eq:5Dlagr}), we will instead look specifically for supersymmetric solutions. A geometry that satisfies the (much easier) supersymmetry equations in a theory will also automatically satisfy the equations of motion.\footnote{This is usually not strictly speaking true, rather, the condition is usually ``supersymmetry equations \emph{plus Bianchi identities} implies equations of motion''; for the sake of simplicity we will not discuss this subtlety.}

It can be shown that the metric and gauge fields of a general supersymmetric solution to (\ref{eq:5Dlagr}) can be parametrized as:\footnote{This is actually only the most general metric for the ``timelike'' class of supersymmetric metrics; there is also the ``null'' class \cite{Gauntlett:2002nw}.} 
\be \label{eq:defds5Theta} ds_5^2 = -Z^{-2} (dt + k)^2 + Z ds_4^2, \qquad \Theta^I = dA^I + d\left( Z_I^{-1}(dt+k)\right),\ee
where we have used the quantities $Z, Z_I$ defined in (\ref{eq:5Dscalars}). The one-form $k$ is the rotation form and parametrizes the time-space cross-terms.
At this point, the expression for $\Theta^I$ is simply a convenient  definition that isolates the ``magnetic'' part of the gauge field strength $dA^I=F^I$.

Supersymmetry further demands that the four-dimensional ``base space'' $ds_4$ is \emph{hyper-K\"ahler}, a stringent mathematical demand. The functions $Z_I$, the one-form $k$, and the two-forms $\Theta^I$ are also restricted to ``live'' on this base space --- so, in particular, they cannot depend on the time $t$. The supersymmetry equations can then be expressed in terms of the quantities $\Theta^I, Z_I, k$ as:\footnote{In particular,  solutions to these supersymmetry equations will preserve 4 supercharges. In the five-dimensional supergravity (\ref{eq:5Dlagr}), which has eight supercharges, they are 1/2 BPS \cite{Gauntlett:2004qy}; in an uplift to the full ten-dimensional string theory (which has 32 supercharges), they are 1/8 BPS solutions. }
\begin{align}
 \label{eq:5DTheta}\Theta^I &= *_4 \Theta^I,\\
 \label{eq:5DZI}\nabla^2 Z_I &= \frac12 C_{IJK} *_4 (\Theta^J\wedge \Theta^K),\\
 \label{eq:5Dk}dk + *_4 dk &= Z_I \Theta^I,
\end{align}
where the Hodge stars $*_4$ are on the four-dimensional base space $ds_4^2$.
Remarkably, this is actually a step-wise linear system of equations; if we solve them in this order, each equation is only a \emph{linear} equation in the unknown functions --- and so can be solved in full generality. Specifically, the path to finding solutions has four steps:
\begin{enumerate}
 \item Find a four-dimensional \emph{hyper-K\"ahler} base space ($ds_4$). This is actually the hardest step, since the equations to restrict to hyper-K\"ahler are non-linear. We will restrict ourselves to a class of hyper-K\"ahler spaces called \emph{Gibbons-Hawking} spaces. (Integration ``constant'': $V$)
 \item Solve (\ref{eq:5DTheta}) for $\Theta^I$.  (Integration ``constants'': $K^I$)
 \item Solve (\ref{eq:5DZI}) for $Z_I$. (Integration ``constants'': $L_I$)
 \item Solve (\ref{eq:5Dk}) for $k$. (Integration ``constant'': $M$)
\end{enumerate}
The last three steps involve integrating linear differential equations. At each of these steps, there will be a choice of ``integration constants'' --- more precisely, arbitrary harmonic functions --- which we will call $V,K^I,L_I,M$. These eight harmonic functions will then completely determine the supersymmetric solution, and choosing them is equivalent with choosing a particular multi-centered geometry. Let us now discuss these four steps in more detail.

\subsubsection{Finding a hyper-K\"ahler base space: Gibbons-Hawking metrics}\label{sec:GHmetrics}
We will choose the four-dimensional hyper-K\"ahler base space $ds^2_4$ to be of the Gibbons-Hawking class. These metrics are not the most general hyper-K\"ahler space (which is not known in full generality), but they are the unique set of hyper-K\"ahler metrics with a tri-holomorphic $U(1)$ isometry.

We can write a Gibbons-Hawking metrics as a $U(1)$ fibration over a flat $\mathbb{R}^3$ base:
\be \label{eq:metricGH} ds_4^2 = V^{-1}(d\psi + A)^2 + V ds_3^2,\ee
where the flat $\mathbb{R}^3$ is simply:
\be ds_3^2 = dx^2 + dy^2 + dz^2 = dr^2 + r^2d\theta^2 + r^2\sin^2\theta d\phi^2,\ee
where $V$ is a harmonic function on this $\mathbb{R}^3$, so that:
\be \nabla^2 V = 0, \ee
and the one-form $A$ is related to $V$ by:
\be \vec{\nabla} \times \vec{A} = \vec{\nabla} V,\ee
where again all quantities are considered as ``living'' on the (flat) $\mathbb{R}^3$. The fourth coordinate $\psi$ is periodic, $\psi \sim \psi + 4\pi$. 

Since $V$ is a harmonic function on $\mathbb{R}^3$, we can write it as:\footnote{Other types of (non-pointlike) sources are also possible in harmonic functions, but are not generically possible in our bubbling solutions (although some exceptions  exist \cite{Bacchini:2021fig}).}
\be\label{eq:exprV} V = v^0 + \sum_{i=1}^n \frac{v^i}{r_i},\ee
where $v_0,v_i$ are constants, and we have introduced $n$ singularities or ``centers'' for $V$, located (in $\mathbb{R}^3$) at positions $\vec{r}_i$; the distances $r_i$ are then given by the usual $\mathbb{R}^3$ distance function:
\be r_i \equiv |\vec{r} - \vec{r}_i|.\ee
Of course, the expression (\ref{eq:exprV}) is reminiscent of other instances where we have harmonic functions on $\mathbb{R}^3$, for example when $V$ represents an electrostatic potential of $n$ point charges.

\subsubsection{Solving for $\Theta^I$}
Once we have specified our four-dimensional base space $ds_4^2$, we can turn to equation (\ref{eq:5DTheta}) which determines the two-forms $\Theta^I$. From the definition  (\ref{eq:defds5Theta}) of $\Theta^I$, it is clear that these forms must be closed, $d\Theta^I=0$. Then, (\ref{eq:5DTheta}) tells us that $\Theta^I$ must be \emph{harmonic, self-dual} forms on $ds_4^2$. Cohomology on Gibbons-Hawking spaces ensures us we can then express these forms as a linear combination of a basis $\Omega_+^a$ (with $a=1,2,3$) of harmonic self-dual forms on $ds_4^2$ (see Section 4.2 in \cite{Bena:2007kg} for an explicit choice of $\Omega_+^a$) as:
\be \Theta^I = \sum_{a=1}^3 \partial_a (V^{-1} K^I) \Omega^a_+,\ee
where $K^I$ are arbitrary harmonic functions on the (flat) $\mathbb{R}^3$. The choice of $K^I$ then completely determines the two-forms $\Theta^I$.

\subsubsection{Solving for $Z_I$}
Once the $\Theta^I$ are determined, (\ref{eq:5DZI}) are linear equations determining the $Z_I$, solved by:
\be Z_I = \frac12 C_{IJK} \frac{K^J K^K}{V} + L_I,\ee
where the $L_I$ are arbitrary harmonic functions on $\mathbb{R}^3$.

\subsubsection{Solving for $k$}
Finally, once $\Theta^I,Z_I$ are completely fixed, (\ref{eq:5Dk}) is a linear equation for the one-form $k$, which is solved by:
\begin{align} k &= \mu(d\psi + A) + \omega,\\
 \mu &\equiv \frac16 C_{IJK} \frac{K^I K^J K^K}{V^2} + \frac{1}{2V} K^I L_I + M,\\
 \label{eq:omega}\vec{\nabla}\times \vec{\omega} &\equiv V \vec{\nabla} M - M \vec{\nabla} V + \frac12 \left( K^I \vec{\nabla} L_I - L_I \vec{\nabla} K^I\right),
\end{align}
where $M$ is again a new, arbitrary harmonic function on $\mathbb{R}^3$. The one-form $\omega$ is determined by (\ref{eq:omega}); note that the integration constants that come from solving this equation are not physically relevant --- they can be absorbed in a shift of the $t$ coordinate.

\subsubsection{Summary: eight harmonic functions}
After following these steps, we now have a complete supersymmetric solution that is determined by eight harmonic functions on the flat $\mathbb{R}^3$, which we can collectively denote as an eight-dimensional ``vector'' $H\equiv (V, K^I, L_I, M)$; with $\nabla^2 H=0$. These harmonic functions have the form:
\be H = h^0 + \sum_{i=1}^n \frac{\Gamma^i}{r_i},\ee
The constant terms $h^0$ are called the asymptotic ``moduli'' of the solution, explicitly:
\be h^0 \equiv (v^0, k_I^0, l_I^0,m^0) = (v^0, k_1^0,k_2^0,k_3^0,l_1^0,l_2^0,l_3^0,m^0).\ee
There are $n$ locations in $\mathbb{R}^3$ where the harmonic functions have singularities --- these are the ``centers'' of the solution. There is no limit on the number $n$ of centers that is allowed in the solution --- although finding a regular solution with many centers will be hard due to the bubble equations  (see below).\footnote{Also, in the full quantum theory, the fluxes (charges) will need to be quantized, so this will further restrict the allowed solutions. We have also not discussed the gauge transformations that relate physically equivalent configurations; see Section 5.2 of \cite{Bena:2007kg} for more information.} The location in $\mathbb{R}^3$ of the $i$-th center is $\vec{r}_i$ and $r_i = |\vec{r} - \vec{r}_i|$ is the (flat) $\mathbb{R}^3$ distance to it. The ``charge'' vector $\Gamma^I$ of the $i$-th center is:
\be \Gamma^i \equiv (v^i, k_I^i, l_I^i, m^i)=\left(v^i, k_1^i, k_2^i, k_3^i, l_1^i, l_2^i, l_3^i, m^i\right).\ee
The harmonic functions $H$, and so also the entire solution, are completely determined by the constants $h^0$ together with the locations $\vec{r}_i$ and charge vectors $\Gamma^i$ of the $n$ centers.

\subsection{Regularity: the bubble equations}\label{sec:bubbleeqs}
We have constructed supersymmetric solutions that depend on eight arbitrarily chosen harmonic functions. However, not all such supersymmetric solutions are physical, and we must impose further constraints to ensure they are regular and well-behaved. One important constraint is the absence of \emph{closed timelike curves} or CTCs. This implies the conditions \cite{Bena:2007kg}:
\be \label{eq:CTCQVZ} \mathcal{Q} \equiv Z_1 Z_2 Z_3 V - \mu^2 V^2 \geq 0, \qquad V Z_I \geq 0,\ee
need to be satisfied \emph{everywhere} (and for each $I$).

Near a center, a necessary (but not sufficient) condition for CTCs to be absent is that $d\psi$ does not become timelike; this leads to the \emph{bubble equations} (of which there is one for each center $i$):
\be \label{eq:bubbleeq} \sum_{j\neq i} \frac{\langle \Gamma^i, \Gamma^j\rangle}{r_{ij}} = \langle h^0, \Gamma^i\rangle,\ee
where we have defined the intercenter distance $r_{ij} \equiv |\vec{r}_i-\vec{r}_j|$ and we define the symplectic product of two charge vectors as:
\be \label{eq:symplprod} \langle \Gamma^i, \Gamma^j\rangle \equiv \left(m^iv^j - \frac12 \sum_I k_I^i l_I^j\right) - (i\leftrightarrow j).\ee
The bubble equations give complicated non-linear relations that the intercenter distances and the center charges need to satisfy. This makes it difficult to find solutions when the number of centers is large. Note that the symplectic product (\ref{eq:symplprod}) naturally pairs electromagnetically dual charges with each other.

If we want a \emph{smooth, horizonless} solution --- meaning one without any horizons or gravitational singularities --- each center's charges must further satisfy:
\be \label{eq:lmi} l_I^i = -\frac12 C_{IJK} \frac{k^i_Jk^i_K}{v^i}, \qquad m^i = \frac{1}{12} C_{IJK} \frac{k^i_Ik^i_Jk^i_K}{(v^i)^2},\ee
so that all of the $L_I$ and $M$ charges are determined by the $V,K^I$ charges.

Other regular, physical solutions exist where the centers do not satisfy (\ref{eq:lmi}). For example, regular supersymmetric multi-centered solutions with black holes exist, where the black hole centers do not satisfy (\ref{eq:lmi}). Also, it is possible to relax (\ref{eq:lmi}) to obtain solutions which have no horizon but include certain ``allowed'' singularities --- these are certain singularities of which we know their nature in string theory, such as brane singularities.

\subsection{Moduli, asymptotics, and charges}\label{sec:moduli}
The constants in the harmonic functions, $h^0 \equiv (v^0, k_I^0, l_I^0,m^0)$, are the \emph{moduli} of the solution. One of these (typically $m_0$) is determined by the sum of the bubble equations (\ref{eq:bubbleeq}), i.e. $\langle h, \sum_i \Gamma^i\rangle = 0$. The choice of the (rest of the) moduli determines what the asymptotic spacetime looks like, and in particular if the solution ``lives'' in four or five dimensions. We will discuss the most common choices below for five- and four-dimensional flat asymptotics. We will not discuss details of other possible choices for moduli that would also give these asymptotics, nor moduli that lead to other possible asymptotics (such as $AdS_2\times S^3$).\footnote{It is not possible to choose the asymptotics to be five-dimensional AdS --- we would need to be working in a gauged supergravity theory  to have such asymptotics. Note that no multi-centered solutions with AdS asymptotics are known; heuristically, it is much harder to construct such solutions since the AdS gravitational potential provides an extra contracting force on any extended object. Even so, multi-centered configurations in AdS may still exist \cite{Monten:2021som}.}

\subsubsection{Five-dimensional asymptotics \texorpdfstring{$\mathbb{R}^{4,1}$}{R41}}
To obtain a five-dimensional solution, we can choose:
\be v^0 = k^0_I = 0, \qquad l^0_I = 1, \qquad \sum_i v^i = 1.\ee
In this five-dimensional spacetime, the geometry carries three electric charges (one for each gauge field $F^I$) as well as two angular momenta $J_L,J_R$.
For solutions that are  smooth and horizonless and so satisfy (\ref{eq:lmi}), the electric charges are given by:
\be
 Q_I = -2 C_{IJK} \sum_i \frac{\tilde k^i_J \tilde k^i_K}{v^i}, \qquad \tilde k^i_I \equiv k^i_I - v^i\sum_j k^j_I,\ee
 and one of the angular momenta is given by:
 \be 
 J_R = \frac43 C_{IJK} \sum_i \frac{\tilde k^i_I \tilde k^i_J \tilde k^i_K}{(v^i)^2}.
\ee
The expression for $J_L$ is more complicated (see eqs. (152)-(154) in \cite{Bena:2007kg}). As mentioned, such five-dimensional horizonless, smooth microstate geometries are microstates of the BMPV black hole with the same electric charges $Q_I$; note that the BMPV black hole has $J_L = 0$.\footnote{Whether microstate geometries with $J_L\neq 0$ can be considered ``microstates'' of the BMPV black hole depends on the particular ensemble in which we consider the black hole, i.e. whether we keep $J_L$ fixed in the ensemble or not. See e.g. \cite{Balasubramanian:2005qu}.}

\subsubsection{Four-dimensional asymptotics \texorpdfstring{$\mathbb{R}^{3,1}\times S^1$}{R31xS1}}
In order to have four-dimensional asymptotics, we must ensure that the $\psi$ circle becomes of constant size at infinity (instead of growing with $r$). One possible choice is:
\be v^0 = 1, \qquad k^0_I = 0, \qquad l^0_I = 1,\ee 
Another possible choice that we will see is:
\be v^0 = l^0_I = 0, \qquad k^0_I = 1, \qquad m^0 = -\frac12.\ee
In general, the condition for having an asymptotically four-dimensional solution is
\be \lim_{r\rightarrow\infty}\mathcal{Q}=1,\ee
where $\mathcal{Q}$ is defined in (\ref{eq:CTCQVZ}) and is called the \emph{quartic invariant}. Note that in all cases, the sum of the bubble equations (\ref{eq:bubbleeq}) must still be satisfied, i.e. $\langle h, \sum_i \Gamma^i\rangle = 0$.

It is possible to dimensionally reduce these asymptotically $\mathbb{R}^{3,1}\times S^1$ solutions over the $\psi$ circle to obtain a solution in a four-dimensional (STU) supergravity theory. We will not discuss the details of this four-dimensional reduced solution (see e.g. appendix A of \cite{Bossard:2014ola}), except to note that the four-dimensional metric is  given by:
\be ds_\text{(4D)}^2 = -\mathcal{Q}^{-1/2} (dt+\omega)^2 + \mathcal{Q}^{1/2} ds_3^2.\ee
Note that a smooth center in five dimensions satisfying (\ref{eq:lmi}) will give a (naked) singularity when dimensionally reduced to four dimensions. This is because such a smooth center is precisely a location where the $\psi$ circle pinches off (i.e. becomes zero size) and so is an ill-behaved point under the dimensional reduction.

\subsection{String theory brane interpretations}\label{sec:braneinterpret}
So far, we have discussed multi-centered solutions as solutions of a five-dimensional supergravity theory. This supergravity theory can be obtained from string theory by a toroidal reduction. More precisely, we can uplift a five-dimensional solution to an 11-dimensional solution of M-theory on $T^2\times T^2\times T^2$ with metric:
\be \label{eq:11Dmetric} ds_{11}^2 = ds_5^2 + \left( \frac{ Z_2 Z_3}{Z_1^2}\right)^{1/3} (dT_1)^2 + \left( \frac{ Z_1 Z_3}{Z_2^2}\right)^{1/3} (dT_2)^2 + \left( \frac{ Z_1 Z_2}{Z_3^2}\right)^{1/3} (dT_3)^2,\ee
where $(dT_I)^2 \equiv dx_I^2 + dy_I^2$ is the flat metric on the $I$-th two-torus.

In this M-theory solution, the five-dimensional electric charges in $L_I$ (and thus $Q_I$) are interpreted as stacks of M2-branes wrapping the tori $T_I$. The magnetic duals of M2 branes are M5 branes, and correspondingly the $K^I$ charges with $I=1,2,3$ (which are the magnetic dual of the $L_I$ charges) come from M5 branes wrapping  $T_2\times T_3, T_1\times T_3$, or $T_1\times T_2$. Note that these M5-branes wrap four compact directions on the tori \emph{and} extend along a \emph {closed curve in the five-dimensional non-compact space} --- so this gives rise to a \emph{dipole} (M5) magnetic charge in five dimensions. This is entirely analogous to the D2-brane dipole we discussed in the context of the supertube transition in Section \ref{sec:supertube}. Finally, the $V$ charges are KK monopole ``charges'' (along $\psi$) and $M$ charges are momentum charges along the $\psi$ direction.

If we dimensionally reduce the 11-dimensional metric (\ref{eq:11Dmetric}) over the $\psi$ circle, we obtain a solution in type IIA supergravity and string theory. In this frame, the $V, K^I,L_I$, and $M$ charges are D6-branes, D4-branes, D2-branes, and D0-branes, respectively. A \emph{smooth} solution satisfying (\ref{eq:lmi}) can be interpreted as D6-branes wrapping all three tori ($T_1\times T_2\times T_3$), with magnetic fluxes on each of the tori that induces the D4-charges --- very much like the magnetic flux on D2-branes in Section \ref{sec:supertube} induces D0-brane charges on it. Then, the D2- and D0-brane charges are induced by ``$F\wedge F$'' and ``$F\wedge F\wedge F$'' Chern-Simons terms involving the (same) magnetic fluxes on the D6-brane worldvolume.

There are many dualities in string theory that allows us to change ``frames'', enabling us also to easily interpret our five-dimensional solutions as different kinds of brane solutions. Another notable frame is one where the three electric charges in five-dimensions become D1, D5, and P (momentum) brane charges.

\subsection{Bubbles as horizon-scale microstructure}\label{sec:horizonscalebubbles}
\begin{figure}[ht]\centering
\includegraphics[width=0.5\textwidth]{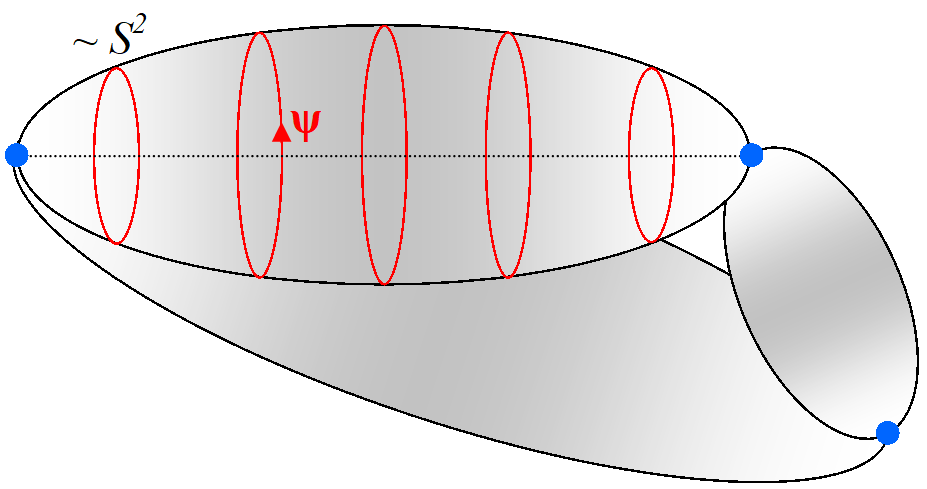}
 \caption{A depiction of the smooth, bubbled microstructure. At each center (blue point), the $\psi$ circle shrinks to zero size; between each pair of centers, there is a topological $S^2$ bubble consisting of the path between the centers (dotted line) and the $\psi$ circle (in red). Each bubble is kept from collapsing due to magnetic flux threading it.}
 \label{fig:bubbles}
\end{figure}

From the Gibbons-Hawking metric (\ref{eq:metricGH}), it is apparent that the singularities of $V$ are where the $\psi$ circle ``pinches off'' and becomes zero size. These singularities of $V$ are precisely the ``centers'' of the solution. If we take an arbitrary path in $\mathbb{R}^3$ from one center to another, then this path together with the $\psi$ circle defines a topological $S^2$ sphere; see Fig. \ref{fig:bubbles}. (Any path between the same two centers will define topologically equivalent objects, so the precise path taken is irrelevant.) Such topological $S^2$ structures are precisely what we call the \emph{bubbles} in this geometry.  Since this reasoning applies to any pair of centers, the total topology of the five-dimensional spacetime will involve some finite product of $S^2$'s \cite{Gibbons:2013tqa}.

An $S^2$ bubble would want to collapse in gravity if not prevented by some mechanism. Here, that mechanism is the \emph{magnetic flux} that threads this bubble. The magnetic flux $\Pi^I_{ij}$ through the bubble between the $i$-th and $j$-th centers is \cite{Bena:2007kg}:
\be \Pi_{ij}^I = \frac{k_j^I}{v_j} - \frac{k_i^I}{v_i}.\ee
As mentioned above, the $K^I$ charges and thus the magnetic fluxes can be interpreted as M2-branes wrapping different tori in an M-theory frame. These M2-branes are not localized --- they are ``branes dissolved in flux'' (see Section \ref{sec:branesflux}) --- and a result, the five-dimensional metric is not singular.

The multi-centered bubbling geometries are an amazing confluence of the concepts we have discussed above. These  geometries are completely smooth and horizonless, but consist of topological ``bubble'' structures that are held stable from collapse by magnetic fluxes, which in turn come from string theoretic branes ``dissolving'' or transitioning into non-singular geometric configurations. These geometries consist of \emph{smooth, stable microstructure}!

Finally, although we did not explicitly discuss this yet, we can also make this microstructure \emph{horizon-scale}. It turns out that we can make this microstructure as compact as we want: in the class of so-called \emph{scaling solutions}, we can essentially have the geometry's bubbles sit arbitrarily far down a redshift throat (i.e. ``being close to the horizon''); see Fig. \ref{fig:scaling}. We will touch on some more details of such solutions in Exercise \ref{ex:scaling} below.

  \begin{figure}[ht]\centering
  \begin{subfigure}[t]{0.22\textwidth}\centering
    \includegraphics[width=\textwidth]{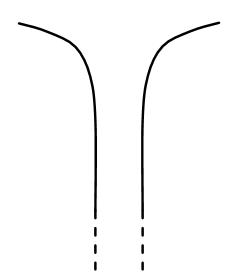}
    \caption{Black hole}
  \end{subfigure}\hfill
  \begin{subfigure}[t]{0.7\textwidth}\centering
    \includegraphics[width=\textwidth]{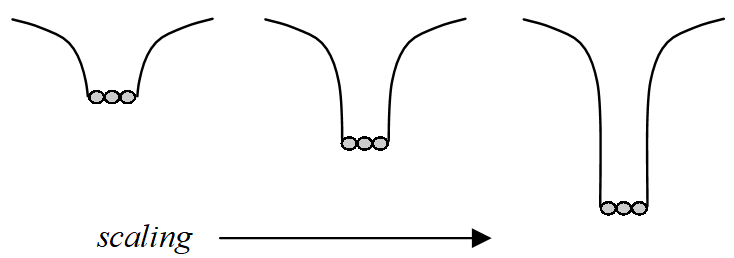}
    \caption{Scaling multi-centered bubbling geometry}
  \end{subfigure}
 \caption{A cartoon of the infinitely deep black hole redshift throat (left), and the finite redshift throats of a series of scaling multi-centered bubbling microstate geometries (right). The black hole redshift throat is infinite since $g_{tt}\rightarrow 0$ as we approach the horizon; the depth of the microstate geometry's redshift throat is always finite ($g_{tt}<0$ everywhere). As we approach the scaling limit, the microstate geometry's throat becomes arbitrarily deep ($|g_{tt}|\ll 1$). Note that the bubbles at the ``end'' of the throat remain a finite, constant \emph{proper} size as the geometry approaches the  scaling limit.}
 \label{fig:scaling}
\end{figure}

\section{Superstrata}\label{sec:SS}

This section will introduce the \emph{superstrata} family of microstate geometries. We will not cover them at the level of detail as Section \ref{sec:multicenter} covered the multi-centered bubbling solutions; rather we will limit the discussion to an emphasis on the conceptual picture of superstrata, and a brief exhibition of (only) the metric in Section \ref{sec:SSsol}.

The main lecture note available to learn more about superstrata is \cite{Shigemori:2020yuo}. The first paper exhibiting superstrata is \cite{Bena:2015bea} and is also very readable. The discussion of Section \ref{sec:stringmomentum} is heavily inspired by \cite{Mathur:2005zp}. The multi-mode superstrata discussed in Section \ref{sec:SSsol} were first found in \cite{Heidmann:2019xrd}; see especially Appendix A therein for a handy summary of the metric and other fields of most of the known (multi-mode) superstrata geometries.

\subsection{Bubbles with fluctuating flux profiles}

The multi-centered bubbling geometries of Section \ref{sec:multicenter} represent an impressive technical achievement and give us a large family of microstate geometries. From a five-dimensional perspective, these are microstates of the five-dimensional BMPV black hole. This black hole has three charges; when these charges are equal, its entropy scales as\footnote{Of course, we are interested in the more generic case when the charges are not equal; it is simply easier to consider equal charges when we are giving rough arguments on how a quantity ``scales'' with the charges.} $S_\text{BMPV}\sim Q^{3/2}$. However, we can use counting arguments to estimate that the multi-centered microstate geometries can only account for (at most) a fraction of the entropy: $S_\text{(bubbles)}\sim Q^{5/4}$ \cite{Bena:2010gg}. (Recall that the number of states $\mathcal{N}$ is related to the entropy as $\mathcal{N}\sim e^S$.) The bubbling geometries can only account for an \emph{exponentially small} fraction of the microstates of the BMPV black hole.

\begin{figure}[ht]\centering
\includegraphics[width=0.9\textwidth]{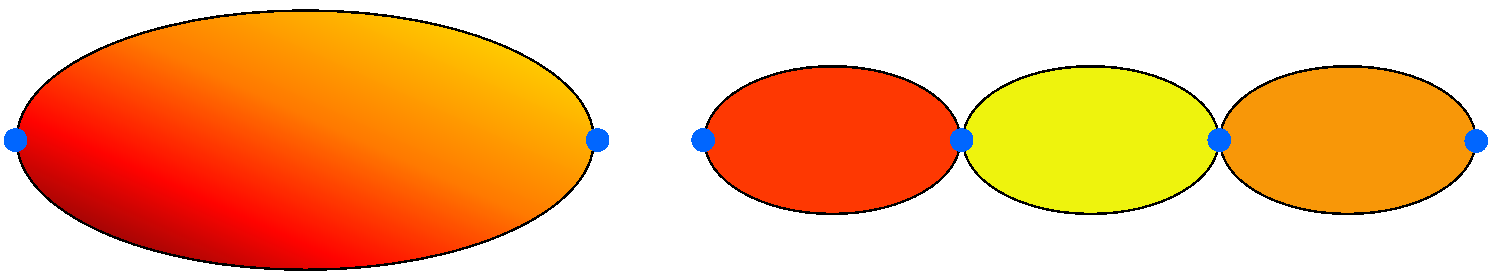}
 \caption{A single bubble with varying flux profile (left), and a configuration of three bubbles where each bubble has rigid, unvarying flux (right). (Centers are indicated as blue points.)}
 \label{fig:rigidvaryingfluxbubble}
\end{figure}

Luckily, we can do better --- with \emph{superstrata}. One way to picture a superstratum is as a geometry with a single bubble (and thus two centers); this bubble must be kept from collapsing by magnetic flux threading it. In the multi-centered solutions we discussed above, this flux was a constant over the entire bubble. In a superstratum, this flux is  allowed to \emph{vary} over the bubble; see Fig. \ref{fig:rigidvaryingfluxbubble}. There are many choices of ``flux profile'' that are possible on the bubble; each one of these profiles gives rise to a different geometry and so a different (micro)state.

To include these non-trivial magnetic flux profiles, yet another (compact spatial) dimension is necessary --- superstrata live in \emph{six} dimensions. It turns out that allowing the flux on a \emph{single} bubble to vary in this way gives rise to exponentially more possible states then allowing for multiple bubbles with a rigid flux profile (as in the multi-centered geometries) \cite{Shigemori:2019orj,Mayerson:2020acj} --- so even single bubble superstrata will give us access to many more microstates of the BMPV black hole than the multi-centered geometries did.

\subsection{Supertube puffing up revisited: strings with momentum}\label{sec:stringmomentum}
Let's now revisit the supertube transition. We discussed this in Section \ref{sec:supertube} from the perspective of F1 and D0 brane charges ``puffing up'' into a D2 (dipole) brane with electromagnetic fluxes living on it. Here, we will instead consider a different system: an (F1) string with (P) momentum running along it. The two systems are related by string dualities, but it is still useful to consider this different perspective.

\begin{figure}[ht]\centering
  \begin{subfigure}[t]{0.5\textwidth}\centering
    \includegraphics[width=\textwidth]{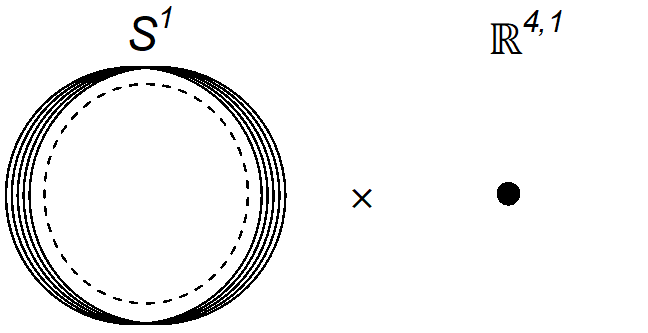}
    \caption{Naive singular geometry}
  \end{subfigure}\hfill
  \begin{subfigure}[t]{0.5\textwidth}\centering
    \includegraphics[width=\textwidth]{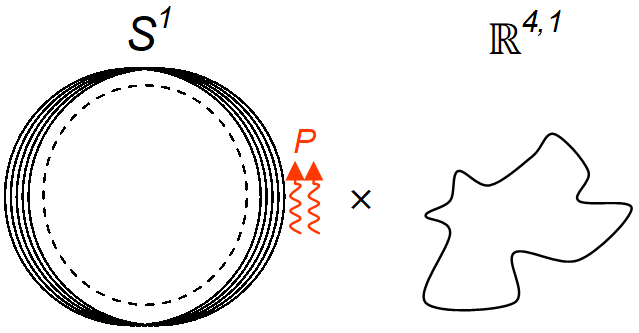}
    \caption{Puffed-up supertube geometry}
  \end{subfigure}
 \caption{In the naive picture, the string wraps a compact $S^1$ direction but is simply a point particle in the non-compact space. However, to carry momentum $P$ along the $S^1$ direction, the string must ``puff up'' into a closed curve in the non-compact space (while remaining wrapped along the $S^1$).}
 \label{fig:luninmathur}
\end{figure}

Consider an F1 string that wraps some compact $S^1$ direction. From the perspective of the other, non-compact directions, this string is simply a point particle; see Fig. \ref{fig:luninmathur}. (For simplicity, we are ignoring any other compact directions besides the $S^1$ that might be present.) Now, we let the string also carry a momentum (P) along this same $S^1$ direction.

This may not seem like an exciting thought exercise, until we remember a basic fact in string theory: \emph{strings cannot carry longitudinal excitations!} In other words, the F1 string \emph{cannot} carry P charge along the direction that it is pointing.

There is only one way that this paradox can be resolved: the string's profile \emph{must change}. Instead of being a point particle in the non-compact directions, it must ``puff up'' into some (contractible) profile in the non-compact space; see Fig. \ref{fig:luninmathur}. By doing so, it can assure that the direction of momentum charge along the $S^1$ direction is always perpendicular to the actual string's direction --- so that the string is only carrying \emph{transverse} excitations, as it should. Note that the resulting configuration still only carries F1 charge along the $S^1$ and P charge along the $S^1$. We have not introduced any additional F1 charge in the system as the string's profile is along a closed, \emph{contractible} cycle in the non-compact space --- this only introduces a (new) \emph{dipole} charge. This is analogous to a current loop in electromagnetism: such a configuration has a magnetic \emph{dipole} charge, but of course no magnetic monopole charge.

We discussed F1 and D0 charges puffing up to create an additional D2 dipole charge in Section \ref{sec:supertube}; here, we see that F1 and P charge puffs up to give an additional F1 dipole charge. These are just a few examples of a general phenomenon: (monopole) charges of a black hole want to ``puff up'' in the black hole's microstate geometries and create additional dipole charges which are not present in the original black hole geometry. In the multi-centered bubbled geometries, these dipole charges were precisely the $k^I_i$'s of the centers.

The F1-P system we discussed here can be dualized using string theory dualities ($S$, $T$, and $U$-dualities, these are called) to the D1-D5 system. In particular, we can consider type IIB string theory in ten dimensions, compactified on a four-torus $T^4$ together with an additional compact circle $S^1$. The non-compact  spacetime is five-dimensional, like for the BMPV black hole. The D1- and D5-branes both wrap the $S^1$ circle, and the D5 branes also wrap the $T^4$. The ``puffing up'' of these D1 and D5 brane charges creates a dipole charge of a geometric Kaluza-Klein monopole (KK). The resulting dipole profile of this KK charge is \emph{arbitrary}; for any such closed profile, we can construct a geometry (smooth and with no CTCs!) --- these are the famous Lunin-Mathur (supertube) geometries \cite{Lunin:2001jy,Lunin:2002iz}. These are completely regular, smooth microstate geometries for the D1-D5 system. By using semi-classical quantization techniques in supergravity, one can quantize these microstate geometries and count them \cite{Rychkov:2005ji}. This counting gives a perfect match with the number of ground states in the dual D1-D5 CFT. In other words, for each of these ground (micro)states, we know precisely what the corresponding supergravity geometry is!

\subsection{Three-charge ``puffing up'': the superstrata}

Although a remarkable achievement, unfortunately the D1-D5 ground states do not correspond to a black hole in supergravity, but rather to a singular D1-D5 geometry where the ``horizon'' has zero size. So we can not quite say that the Lunin-Mathur geometries provide ``black hole'' microstates, but it certainly feels like we are getting close. To construct a system that corresponds to a non-singular black hole of finite size, we can add momentum $P$ along the $S^1$ that the D1 and D5 branes share. This three-charge D1-D5-P system is precisely the one studied by Strominger and Vafa, where they were able to count the microstates in the D1-D5 CFT and found agreement with the entropy of this black hole; we discussed this in Section \ref{sec:SV}.

D1- and D5-branes can ``puff up'' to create KK dipole charges, as mentioned; this dipole charge can have an arbitrary profile. But when we introduce a third charge, additional ``puffing up'' transitions will take place \cite{Bena:2011uw} --- essentially, any two pairs of charges can create new dipole charges by a similar transition. In the end, instead of a one-dimensional profile that we associate to the supertube transition, we expect that a system of three charges will ``puff up'' to create a \emph{two-dimensional} profile or surface --- the \emph{superstratum}.

For the D1-D5 Lunin-Mathur supertube, the one-dimensional (dipole) profile in five spacetime dimensions can be parametrized by four arbitrary functions of one variable $f^j(v)$, where $j=1,\cdots,4$ runs over the spatial non-compact dimensions. All of the $f^j(v)$ are periodic as $f^j(v+L)\sim f^j(v)$ (where $L$ is the parameter length of the string), so that the profile defines a closed loop in the non-compact space. We can expand these real functions in Fourier coefficients $c_n^j$ over the basis of periodic functions over $v$:
\be f^j(v) = \sum_{n>0}\left( c_n^j e^{2\pi i n v/L}+(c_n^j)^* e^{-2\pi i n v/L}\right).\ee
The geometry for \emph{any} such profile functions $f^j(v)$ is known \cite{Lunin:2001jy,Lunin:2002iz}.

For the D1-D5-P superstrata, we instead expect that the most general superstrata can be parametrized by \emph{two} arbitrary holomorphic functions $G_1,G_2$ of \emph{three} complex variables $\xi,\chi,\eta$.\footnote{It is not obvious how to arrive at this counting. Essentially, these three complex variables can be seen as running over the three $U(1)$ isometries of $AdS_3\times S^3$ \cite{Heidmann:2019zws,Heidmann:2019xrd}. Also, note that $|\xi|^2+|\chi|^2+|\eta|^2=1$.} These functions can be expanded in Fourier coefficients as:
\be G_1(\xi,\chi,\eta) = \sum_{k,m,n} b_{k,m,n}\xi^n \chi^{k-m}\eta^m, \qquad G_2(\xi,\chi,\eta) = \sum_{k,m,n} c_{k,m,n}\xi^n\chi^{k-m}\eta^m.\ee
The arbitrary \emph{single-mode} geometry, where only one of the $b_{k,m,n}$ or $c_{k,m,n}$ coefficients are non-zero, is explicitly known \cite{Bena:2015bea,Ceplak:2018pws}. In principle, the procedure is known how to ``superpose'' such single-mode geometries in order to find two- and multi-mode geometries \cite{Heidmann:2019zws}, but in practice the resulting (differential) equations that need to be solved can be prohibitively complicated. Only a few multi-mode geometries have been explicitly constructed \cite{Heidmann:2019xrd}, for example the superposition of $(1,0,n)$ modes (i.e. with an arbitrary number of different $n$'s).

It is interesting to note that for all superstrata, the holographic dual state (in the D1-D5 CFT) is explicitly known \cite{Shigemori:2020yuo,Giusto:2015dfa}. This has allowed for many advances and insights into precision holography of these states. By contrast, the CFT dual or interpretation of a multi-centered geometry of more than two centers is generally unknown.

\subsection{The solutions}\label{sec:SSsol}
We will not discuss the derivation of the superstrata geometries here; for details, see e.g. \cite{Shigemori:2020yuo}, especially Sections 3 and 4.3 therein. We will only briefly discuss the actual solutions, and then only the metric. We will follow the notation of \cite{Heidmann:2019xrd}; see also Appendix A  therein for more details on the other fields in the solution.

As discussed above, superstrata are most naturally constructed in \emph{six}-dimensional supergravity; it is usually not possible to dimensionally reduce them to five dimensions as they have a non-trivial ``profile'' along this sixth dimension. When embedded into flat space \cite{Bena:2017xbt}, they do have the correct \emph{five}-dimensional flat asymptotics, so $\mathbb{R}^{4,1}\times S^1$. We will only discuss the slightly easier case of $AdS_3\times S^3$ asymptotics.
The relevant six-dimensional supergravity theory contains, besides the metric, also a scalar and two three-form gauge field strengths.

The six-dimensional metric is given by:
\be ds_6^2 = -\frac{2}{\sqrt{\mathcal{P}}} (dv+\beta) \left[ du + \omega + \frac{\mathcal{F}}{2}(dv+\beta)\right] + \sqrt{\mathcal{P}}ds_4^2,\ee
where $ds_4^2$ is a flat four-dimensional $\mathbb{R}^4$ base, written in ``spherical bipolar coordinates'' as:
\be ds_4^2 = \Sigma\left( \frac{dr^2}{r^2+a^2} +d\theta^2\right) +(r^2+a^2)\sin^2\theta d\varphi_1^2 + r^2\cos^2\theta d\varphi_2^2,\qquad \Sigma = r^2+a^2\cos^2\theta.\ee
The one-form $\beta$ is:
\be \beta = \frac{R_y a^2}{\sqrt{2}\Sigma} (\sin^2\theta d\varphi_1 -\cos^2\theta d\varphi_2).\ee

\subsubsection{The two-charge round supertube}
A configuration that only carries D1 and D5-brane charges (so no $P$ charge) is the \emph{round supertube}, for which:
\be \mathcal{P} = \frac{Q_1 Q_5}{\Sigma^2}, \qquad \mathcal{F} = 0, \qquad \omega = \omega_0 \equiv \frac{a^2 R_y}{\sqrt{2}\Sigma} (\sin^2\theta d\varphi_1 + \cos^2\theta d\varphi_2).\ee
Regularity of the metric fixes $a$ as:
\be Q_1 Q_5 = R_y^2 a^2.\ee
The D1 (resp. D5) charge of the solution is $Q_1$ (resp. $Q_5$), and the five-dimensional angular momenta are:
\be J_L = J_R = \frac{R_y}{2}a^2.\ee
This two-charge geometry is a special case of the Lunin-Mathur D1-D5 supertube --- in particular, it corresponds to the simple case of a  circular profile for the supertube in the non-compact directions. (Incidentally, this particular simple circular geometry \emph{can} be dimensionally reduced to a two-centered bubbling geometry in five-dimensional supergravity \cite{Raeymaekers:2008gk}.) This round supertube geometry is the ``basis'' upon which the superstrata's ``momentum wave profiles'' are constructed.\footnote{This begs the question: what about superstrata that are constructed on a different ``basis'', i.e. taking a different starting profile from the round supertube? Constructing superstrata in this way is largely an open question.}

\subsubsection{Complex variables and warp factor}
It is convenient to use the following complex variables:
\be \xi = \frac{r}{\sqrt{r^2+a^2}} e^{i \frac{\sqrt{2}v}{R_y}}, \quad \chi = \frac{a}{\sqrt{r^2+a^2}}\sin\theta e^{i\varphi_1}, \quad \eta = \frac{a}{\sqrt{r^2+a^2}}\cos\theta e^{i \left( \frac{\sqrt{2}v}{R_y}-\varphi_2\right)},\ee
which satisfy $|\xi|^2 + |\chi|^2 + |\eta|^2 = 1$. Spatial infinity $r\rightarrow \infty$ corresponds to $\chi=\eta=0$ and $|\xi|=1$. As discussed above, the most general superstrata is expected to depend on two arbitrary functions of these three variables:
\be G_1(\xi, \chi, \eta) = \sum_{k,m,n} b_{k,m,n} \xi^n \chi^{k-m} \eta^m, \qquad G_2(\xi,\chi,\eta) = \sum_{k,m,n} c_{k,m,n} \xi^n \chi^{k-m}\eta^m,\ee
with Fourier coefficients $b_{k,m,n}, c_{k,m,n}$, and where $G_1$ are the ``original'' superstrata modes and $G_2$ are called the ``supercharged'' superstrata. The warp factor is given by:
\be \mathcal{P} = \frac{1}{\Sigma^2}\left( Q_1 Q_5 - \frac{R_y^2}{2} |G_1|^2\right).\ee

\subsubsection{\texorpdfstring{$(1,0,n)$}{(1,0,n)} multi-mode superstrata}
As a concrete example, let's give all of the functions for the $(1,0,n)$ multi-mode superstrata geometries.

The metric for the general multi-mode $(1,0,n)$ superstrata is:
\be G_1(\xi,\chi,\eta) = \chi F(\xi),\ee
where $F(\xi)$ is an arbitrary holomorphic function:
\be F(\xi) = \sum_{n>0} b_n \xi^n,\ee
and the warp factor is:
\be \mathcal{P} = \frac{Q_1Q_5}{\Sigma^2} - \frac{a^2R_y^2}{2(r^2+a^2)\Sigma^2} |F|^2\sin^2\theta.\ee
We can denote the asymptotic (spatial infinity) limit of $F$ as:
\be F_\infty(v) = \lim_{|\xi|\rightarrow 1} F(\xi) = \lim_{r\rightarrow\infty} F(\xi) = F\left(e^{i\frac{\sqrt{2}v}{R_y}}\right),\ee
and then we can express the rest of the metric functions as:
\begin{align}
 \mathcal{F} &= \frac{1}{a^2}( |F|^2 - |F_\infty|^2),\\
 \omega &= \left( 1 - \frac{1}{2a^2}(|F_\infty|^2-c)\right)\omega_0  + \frac{R_y}{\sqrt{2}\Sigma}( |F_\infty|^2 - |F|^2)\sin^2\theta d\varphi_1,
 \end{align}
The constants $a,c$ must satisfy the regularity conditions:
\be c = 2\left( \frac{Q_1Q_5}{R_y^2} - a^2\right) = \frac{1}{\sqrt{2}\pi R_y} \int_0^{\sqrt{2}\pi R_y} dv' |F_\infty(v')|^2 = \sum_{n=1}^\infty |b_n|^2.\ee
With the above expressions, the geometry is completely regular for any holomorphic function $F$ (with $F(0)=0$); in addition, $\mathcal{P}>0$ everywhere and there are no CTCs anywhere \cite{Heidmann:2019xrd}. The solution has D1, D5, $J_L, J_R$ charges just as the round supertube above, and in addition has also a momentum charge of:
\be Q_P = \frac{1}{4\sqrt{2}\pi R_y} \int_0^{\sqrt{2}\pi R_y} dv\left( \xi_\infty F'_\infty \overline{F}_\infty  + \overline{\xi}_\infty F_\infty \overline{F}'_\infty\right) = \frac12 \sum_{n=1}^\infty n |b_n|^2.\ee
Note that the ``single-mode'' superstrata that are typically studied in the literature are those where all $b_n$ coefficients vanish except for one $n>0$.

Even though the general superstrata live in six dimensions, it was found in \cite{Mayerson:2020tcl} that the $(1,0,n)$ superstrata family (and a few others) can be consistently dimensionally reduced along the compact $S^3$ to obtain three-dimensional asymptotically $AdS_3$ solutions.

\section{Microstructure in Observations}\label{sec:obs}

We have now discussed the two most established families of microstate geometries --- multi-centered bubbling solutions and  superstrata. Rather than discussing more microstate geometries in the remainder of these lectures, we will instead briefly discuss a rather new and exciting application of these geometries: using them as a top-down model of horizon-scale microstructure to understand possible quantum gravity effects in observations.

First, we will introduce the concept of \emph{fuzzball phenomenology} and argue for its necessity within gravitational phenomenology of precision black hole observations. Then, in the last two sections, we will briefly touch on the two main classes of precision black hole observations: gravitational waves from black hole mergers, and (electromagnetic) black hole imaging. For each of these topics, we will also briefly touch on a few relevant observables which could distinguish compact objects such as fuzzballs from black holes.

This will be a lightning overview of the topic; for further details, see \cite{Mayerson:2020tpn} (and \cite{Bacchini:2021fig} for fuzzball imaging of Section \ref{sec:FBimaging}). Some interesting (non-fuzzball-centric) reviews on gravitational phenomenology are \cite{Barack:2018yly}, and for ECOs in particular \cite{Cardoso:2019rvt}.

We will also not discuss details of the actual experiments involved. These are (for gravitational waves) in particular the current aLIGO and aVIRGO detectors \cite{Abbott_2009}, the future space-based LISA mission \cite{Danzmann_1996}, and second-generation ground-based experiments such as the Einstein Telescope and the Cosmic Explorer \cite{Kalogera:2021bya}; and (for black hole imaging) the Event Horizon Telescope (EHT) \cite{EHT2019a} and possible future improvements in the next generation EHT (ngEHT) \cite{Blackburn:2019bly} or space-based Very Long Baseline Interferometry (VLBI) \cite{Haworth:2019urs,Gralla:2020srx}.

\subsection{From observations to fuzzball phenomenology}\label{sec:obstofuzzpheno}
The advent of precision black hole observations has opened an unprecedented new window to the universe. Some 400 years ago, the first use of optical telescopes in astronomy opened our eyes to the universe and its electromagnetic signatures. This led to many insights into the nature of the universe and its occupants --- stars, black holes, white dwarves, and so on. It is no exaggeration to compare the recent detection of the first gravitational waves to this advent of optical telescopes; we now have access to a whole new observational window on the universe by an entirely different force of nature. Centuries from now, we will probably still be figuring out the implications of the new physics we are able to observe!

Gravitational wave observations, as well as the recent developments in black hole imaging, probe the strongest gravity regime we have ever been able to observe --- the near-horizon region of black holes. This is an exciting opportunity to see quantum gravity effects at work. \emph{Can we see effects of quantum gravity in these precision black hole observations? If so, what phenomena should we look for to decisively distinguish quantum gravity effects (from general relativity)?} These are the basic questions of gravitational phenomenology.

We discussed arguments that suggest black hole physics can drastically change at the horizon scale in Section \ref{sec:FBparadigm}. Motivated by these arguments, a slightly more concrete question to ask in phenomenology is: \emph{What happens in observations if we replace a (Kerr) black hole by a horizonless, compact object?}

\begin{figure}[ht]\centering
\includegraphics[width=0.3\textwidth]{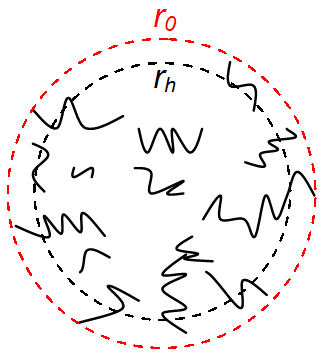}
 \caption{A depiction of a horizonless, exotic compact object (ECO), with the would-be horizon scale $r_h$ and the ECO scale $r_0$ (in red) indicated. The ECO's (smooth) structure extends to $r_0$.}
 \label{fig:ECOscales}
\end{figure}

Such a compact object need not be a fuzzball, but should be approximately the same ``size'' as the black hole --- this means it ``mimics'' the black hole far away from the horizon scale, and only deviates from the black hole geometry approximately at the horizon scale. Such objects are typically referred to as \emph{exotic compact objects} or ECOs \cite{Cardoso:2017njb,Cardoso:2017cqb,Cardoso:2019rvt,Maggio:2021ans}. Their ``compactness'' can be parametrized by a dimensionless parameter $\epsilon$:
\be \label{eq:ECOepsilon} r_0 = r_h(1+\epsilon),\ee
where $r_h$ is the horizon radius of the corresponding black hole (that the ECO replaces), and $r_0$ is the scale at which the ECO geometry starts deviating significantly from the black hole geometry; see Fig. \ref{fig:ECOscales}. We expect the object to be ultra-compact, so $\epsilon\ll 1$, if the object is to mimic a black hole very well. The ``exotic'' in ECO emphasizes that this is quite hard to achieve and requires some kind of ``exotic'' mechanism: Buchdahl's theorem tells us that under reasonable assumptions in general relativity, $\epsilon\geq 1/8$.

\begin{figure}[ht]\centering
\includegraphics[width=0.7\textwidth]{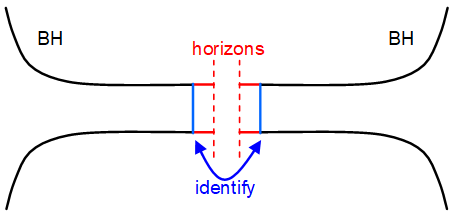}
 \caption{A Solodhukin wormhole, constructed by taking two copies of a black hole geometry, excising their horizon regions (in red) and identifying surfaces outside the horizon (in blue).}
 \label{fig:wormhole}
\end{figure}

There are many different ECOs that are used in gravitational phenomenology. Many are constructed in a bottom-up, effective field theory mindset --- they parametrize or create deviations from general relativity by adding new terms to the action or new objects in the theory. For example, (Solodhukin-type) wormholes \cite{Solodukhin:2005qy,Bueno:2017hyj,Cardoso:2019rvt} are horizonless objects that are constructed by altering a (Schwarzschild or Kerr) black hole geometry by hand; essentially, right before the horizon is reached, a second copy of the asymptotically flat geometry is glued to the original one, creating a wormhole between the two flat geometries --- see Fig. \ref{fig:wormhole}. This gluing procedure in the metric is done ``by hand'', so the resulting geometry is not a solution to any known theory with matter. These geometries can be used to study aspects of horizonless objects that do not require any dynamics of the object itself, since those dynamics are --- by construction --- unknown.

Another example of ECOs are boson stars \cite{PhysRev.172.1331,PhysRev.187.1767,Cardoso:2019rvt}. There are many flavors of such objects, but the most basic example is constructed by adding a minimally coupled scalar field with some potential to the Einstein gravity Lagrangian:
\be \mathcal{L} = R - \frac12 (\partial \phi)^2 - V(\phi).\ee
Boson stars are self-gravitating, compact horizonless solutions in such a theory. These solutions do typically suffer from a trade off between stability and compactness: if the object becomes too compact, it will want to collapse on itself into a black hole.

Instead of constructing ECO models bottom-up, a complementary approach is to consider \emph{top-down} models --- objects which come from a theory of quantum gravity, and so are by construction guaranteed to be consistent, stable objects. Fuzzballs, and in particular microstate geometries, are precisely such objects. They are horizonless objects that can be made arbitrarily compact (recall the scaling solutions mentioned in Section \ref{sec:horizonscalebubbles} and Exercise \ref{ex:scaling}) without imploding on themselves.

Microstate geometries have many limitations, as we discussed in Section \ref{sec:limitations}. Most notably, we do not have any such geometries that can correspond to \emph{realistic} (Kerr) black holes! Accordingly, the philosophy to keep in mind is not that we are looking for direct, specific candidates to replace a given astrophysical black hole. Rather, we study our available microstate geometries as \emph{models of universal mechanisms of horizon-scale microstructure in string theory}. Fuzzballs and microstate geometries provide us with concrete, horizonless objects with microstructure that is supported by consistent quantum gravity mechanisms --- specifically, topological bubbles with flux that live in the additional compact dimensions of string theory. \emph{Fuzzball phenomenology} is then the study of top-down fuzzball geometries as string theory models of quantum gravity effects at the horizon scale for current and future precision black hole observations.

\subsection{Gravitational waves}

Black holes can be captured in each other's orbit. In such a system, they emit gravitational waves as they orbit around each other; these waves carry off energy from the system, so that the black holes spiral closer and closer to each other until they plunge into each other in a violent merger. The merged black hole then emits some final gravitational radiation as it relaxes to a new, steady state.

\begin{figure}[ht]\centering
 \includegraphics[width=0.7\textwidth]{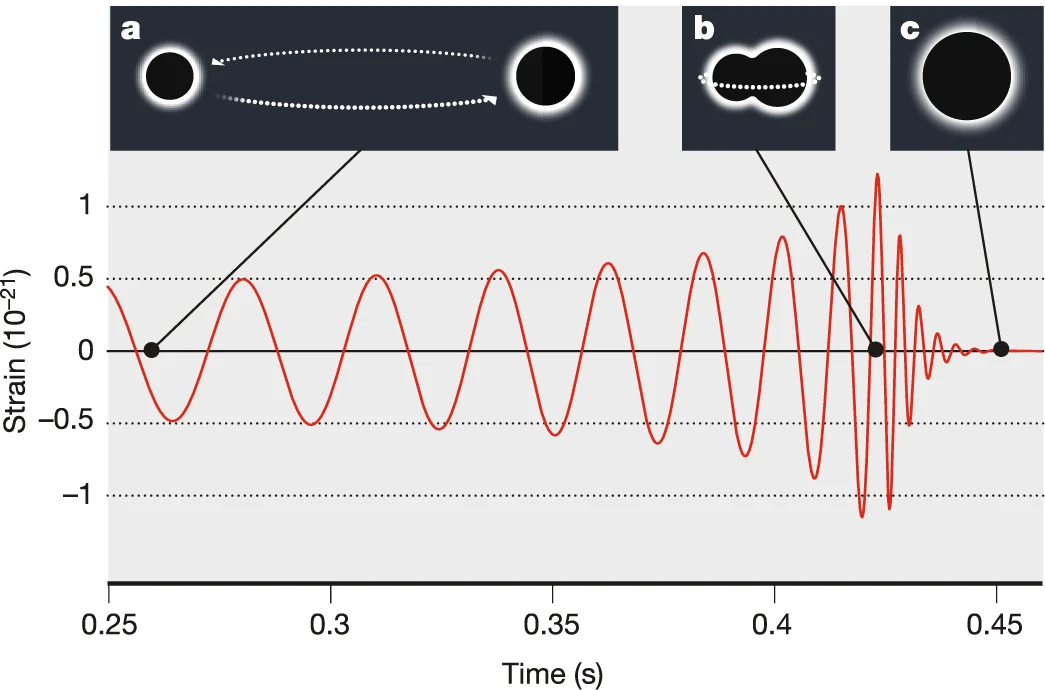}
 \caption[Gravitational wave signal]{The three phases of the merger of a binary system, and a graph of the strength of the gravitational wave emitted in the process. The three phases are (a) inspiral; (b) merger; (c) ringdown.\\
 (This is Fig. 1 of Fig. \cite{GWdawn}, which is adapted from Fig. 2 of \cite{Abbott:2016blz}, licensed under \href{https://creativecommons.org/licenses/by/3.0/}{CC BY 3.0}.)
 }
 \label{fig:GWphases}
\end{figure}

We can divide this process into three distinct phases, see Fig. \ref{fig:GWphases}. First, we have the \emph{inspiral} phase, where the black holes are still relatively far apart from each other. There are a range of perturbative general relativity techniques that can model this phase of the process, for example by using a so-called post-Newtonian expansion where the general relativistic corrections to Newtonian gravitational physics are taken into account order by order (typically in a parameter $\sim 1/c^2$). This phase can also last very long. For example, in extreme mass ratio inspirals (EMRIs), a supermassive black hole captures a smaller object; the smaller object can orbit $\sim 10^5$ times or more before around the supermassive black hole before it plunges in the horizon \cite{Babak:2017tow}.

When the horizons start coming close enough to start overlapping, and form a new, conglomerate horizon, we enter the \emph{merger} phase. This phase is very short (compared to the inspiral) and involves strong gravitational fields --- typically, computationally expensive numerical general relativity simulations are needed to model this phase.

Once the final object has formed, the final black hole enters the \emph{ringdown} phase where it relaxes to a steady state. Perturbative techniques (e.g. adding perturbations on a single, stationary black hole background) are again well suited to model this phase.

The gravitational waves emitted during all three phases of this whole process are sensitive to the details of the black holes involved. In particular, quantum gravity horizon-scale physics may give deviating wave signatures to the classical general relativistic expectation.

We will briefly discuss a few interesting observables \cite{Barack:2018yly} in the inspiral and ringdown phase below, always keeping the application to fuzzballs in mind. It is harder to model possible quantum gravity effects --- especially coming from horizon-scale microstructure --- in the merger phase, due to its strong gravity nature. Of course, the strong gravity of this phase is also precisely why we might expect such quantum effects to be most pronounced in this phase. There are a few hints of possible quantum effects in the merger phase, see e.g. \cite{Hertog:2017vod}.

\subsubsection{Inspiral observables}
As the two black holes orbit each other, they each source a gravitational field that pulls and pushes on the other object. The objects' structure --- especially of or at the horizon scale --- is important to determine the gravitational field that they source and to calculate the \emph{tidal forces} that they exert on each other.

\begin{figure}[ht]\centering
  \begin{subfigure}[t]{0.31\textwidth}\centering
    \includegraphics[width=\textwidth]{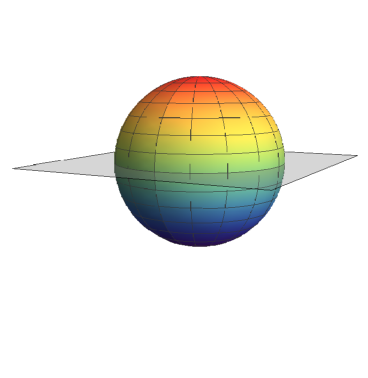}
    \caption{Purely spherical object}
  \end{subfigure}\hspace{1em}
  \begin{subfigure}[t]{0.31\textwidth}\centering
    \includegraphics[width=\textwidth]{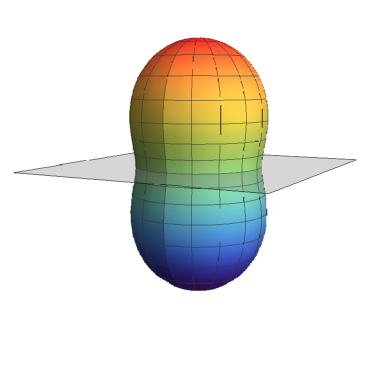}
    \caption{$M_2$ deformation, which preserves equatorial symmetry}
  \end{subfigure}\hspace{1em}
  \begin{subfigure}[t]{0.31\textwidth}\centering
    \includegraphics[width=\textwidth]{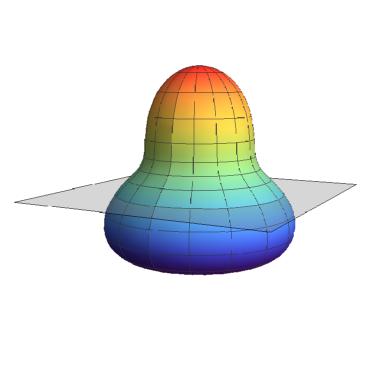}
    \caption{$M_3$ deformation, which breaks equatorial symmetry}
  \end{subfigure}\hfill
 \caption{A spherical object and two deformations: one that preserves equatorial symmetry and one that does not. The equator is indicated as the gray plane. All three objects are axisymmetric.}
 \label{fig:multipoledeform}
\end{figure}

The \textbf{gravitational multipoles} of an object essentially encode its internal gravitational structure; in other words, the multipoles describe the ``bumps'' of an object; see Fig. \ref{fig:multipoledeform}. Gravitational multipoles are analogous to electromagnetic ones, which can be obtained by a multipole expansion of the electrostatic potential:
\be V_\text{EM} = \frac{Q}{r} + \frac{D}{r^2}\cos\theta + \mathcal{O}(1/r^3) .\ee
The leading order term, the monopole, gives the total charge $Q$ in the system. At the next order, we can read off the dipole charge $D$, and so on. For an object in general relativity, one must choose appropriate coordinates which allow reading off two sets of multipoles from the metric:
\be g_{tt} \sim \sum_n \frac{M_n}{r^{n+1}}, \qquad g_{t\phi} \sim \sum_n \frac{S_n}{r^{n+1}}.\ee
(Note that this is only schematic, and assumes axisymmetry; the generalization to non-axisymmetric spacetimes is straightforward \cite{Bena:2020uup,Bianchi:2020miz}.)
The $M_n$ coefficients are called the \emph{mass multipoles} and the $S_n$ are the \emph{current multipoles} (or angular momentum multipoles). The first non-zero multipoles are $M_0=M$, which is the total mass of the object, and $S_1=J$, the angular momentum of the object.

The Kerr black hole in general relativity is entirely determined by its mass $M$ and angular momentum $J=Ma$; its multipole structure can be expressed in terms of $M$ and $a$:
\be \label{eq:Kerrmultipoles} M_{2n} = M(-a^2)^n, \qquad S_{2n+1} = Ma (-a^2)^n,\qquad M_{2n+1} = S_{2n} = 0,\ee
for all $n\geq 0$. Since general relativity is unforgiving in demanding black hole uniqueness, measuring any deviation from this multipole structure would be a clear signal of quantum gravity effects in black hole physics. For example, once the mass $M$ and spin $J$ are measured, any deviation from the Kerr quadrupole value $M_2 = - J^2/M$ would be a smoking gun of deviations from general relativity. It is expected that the future space-based gravitational wave observer LISA will be able to measure any deviations of the dimensionless ratio $M_2/M^3$ from the Kerr value to within one part in $10^4$ \cite{Barack:2006pq}.

Another striking feature of the Kerr multipoles (\ref{eq:Kerrmultipoles}) is that the \emph{odd parity} multipoles $M_{2n+1}$ and $S_{2n}$ all vanish. This is because the Kerr black hole enjoys an \emph{equatorial symmetry}: this is reflection symmetry over its equatorial plane ($z\leftrightarrow -z$ or $\theta\leftrightarrow \pi-\theta$ in usual Cartesian or spherical coordinates); see also Fig. \ref{fig:multipoledeform}. This symmetry is ``accidental'' in that there is no underlying principle in general relativity that ensures its presence. This is in contrast to axisymmetry (i.e. rotational symmetry around the axis of rotation of the black hole), which can be shown to be a consequence of stationarity for Kerr \cite{Townsend:1997ku}. Indeed, many string theory black holes that generalize Kerr break this equatorial symmetry. Measuring a non-zero odd parity multipole $S_{2n}$ or $M_{2n+1}$ would then also be a smoking gun that points towards physics beyond general relativity! EMRIs as seen by LISA are also expected to be able to measure such signals rather precisely; it is estimated that a non-zero value for e.g. $S_2/M^3\gtrsim 10^{-2}$ would be detectable in such processes \cite{Fransen:2022jtw}.

An ECO will typically have different multipoles than the black hole it ``replaces''. The multipoles of supersymmetric multi-centered microstate geometries were discussed in \cite{Bena:2020see,Bianchi:2020bxa,Bena:2020uup,Bianchi:2020miz}; see also \cite{Bah:2021jno} for a discussion of the multipole moments of the so-called almost-BPS microstate geometries. Generically, the scaling parameters for these geometries, which determines (in a heuristic sense) ``how far from the horizon'' the microstructure is located, also sets the scale of how much the microstate geometry's multipoles differ from its corresponding black hole values. The closer to the horizon that the microstructure sits, the smaller the deviations of the multipole moments from the black hole values.

The multipole moments of an object can \emph{change} when a second object is gravitationally pulling on it. How much an object's multipoles change is its \emph{tidal deformability}, which can be parametrized by so-called \textbf{tidal Love numbers} (TLNs) \cite{Kol:2011vg,Cardoso:2017cfl,Pani:2019cyc}. It was recently established that Kerr black holes have \emph{vanishing TLNs} (a rather non-trivial statement \cite{Chia:2020yla,Charalambous:2021mea}!), but in general ECOs do not have vanishing TLNs \cite{Cardoso:2017cfl}. These TLNs leave their imprint on the observed gravitational wave of an inspiralling binary system, so can also be observational smoking guns of horizon-scale microstructure. It is hard to calculate TLNs as it involves solving full linear perturbations (at zero frequency) of a solution in a given gravity theory. Calculating the TLNs of any microstate geometry is still an open problem.

\subsubsection{Ringdown observables}

The ringdown phase can be modelled as a perturbation on top of an otherwise stationary black hole background. The perturbation relaxes exponentially fast as most of the excitation either escapes to spatial infinity or falls into the horizon. This initial relaxation --- called the \emph{initial ringdown} or \emph{prompt ringdown} --- is determined by the \textbf{quasinormal modes} (QNMs) of the black hole \cite{Cardoso:2016rao}. These QNMs are very sensitive to the object's details and ECOs have very different QNMs from black holes \cite{Cardoso:2019rvt}. A precision measurement of the QNMs in the initial ringdown could then be a good probe of the horizon structure. QNMs of multi-centered bubbling microstate geometries were explored in \cite{Ikeda:2021uvc}.

\begin{figure}[ht]\centering
 \includegraphics[width=0.5\textwidth]{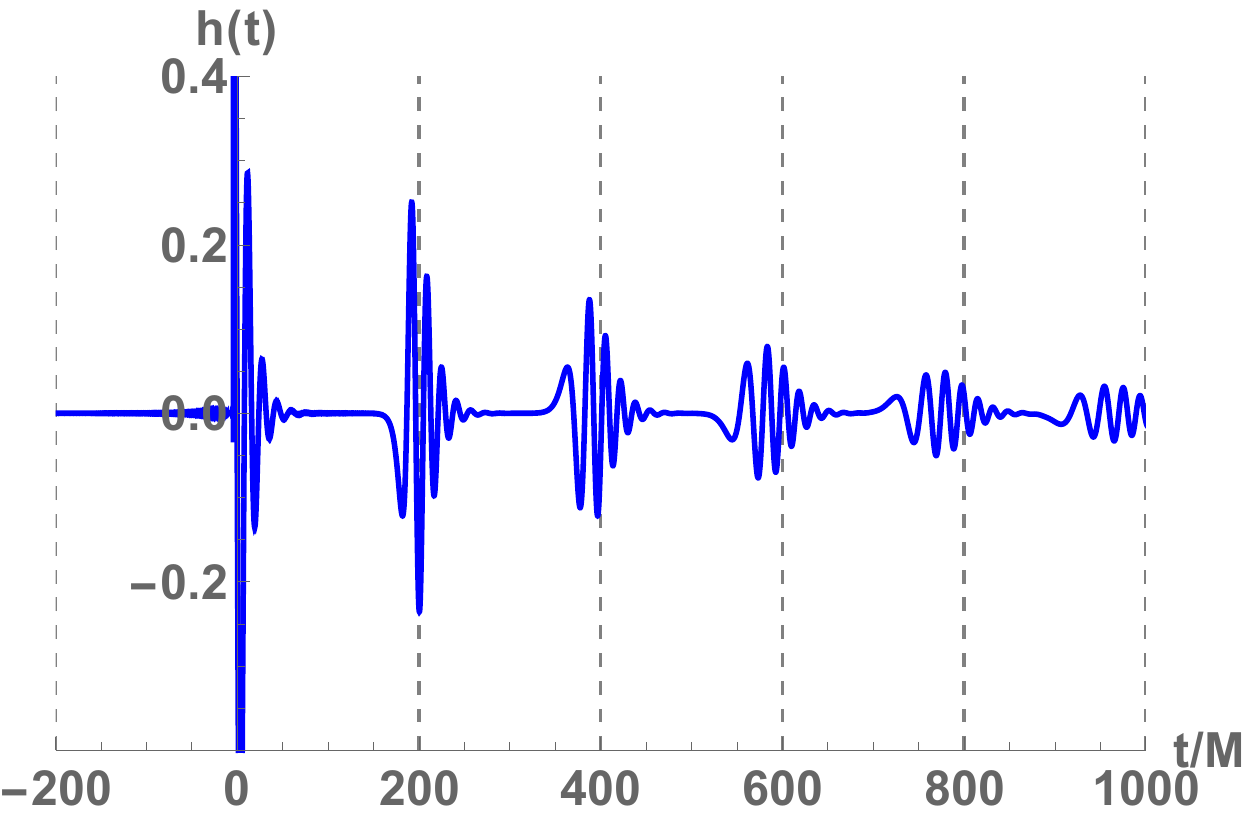}
 \caption[Ringdown echoes]{The (initial) ringdown at $t\sim 0$ in a gravitational wave signal $h(t)$, followed by quasi-periodic and slowly attenuating echoes in the late ringdown signal. The period of the echoes is twice the cavity length, which is taken to be $L = 100M$.\\ (Figure created from data using the analytic template for echoes in non-spinning ECOs \cite{Testa:2018bzd,GWsite,testathesis}; in particular, the template with $L=d=100M, \mathcal{R}=0.75$, grav. polar $\ell=2$ is used.)}
 \label{fig:echoes}
\end{figure}

After the initial ringdown, the perturbation in a black hole spacetime has completely disappeared --- either escaped to infinity or absorbed by the horizon. The black hole has achieved a stationary state, and there is no further gravitational waves being emitted. By contrast, any \emph{horizonless} object will \emph{not} absorb the perturbation wave, but rather can trap a part of it ``inside'' its structure for long periods of time and emit it back out in quasi-periodic, long lives \textbf{echoes} \cite{Cardoso:2017cqb}; see Fig. \ref{fig:echoes}.

\begin{figure}[ht]\centering
 \includegraphics[width=0.6\textwidth]{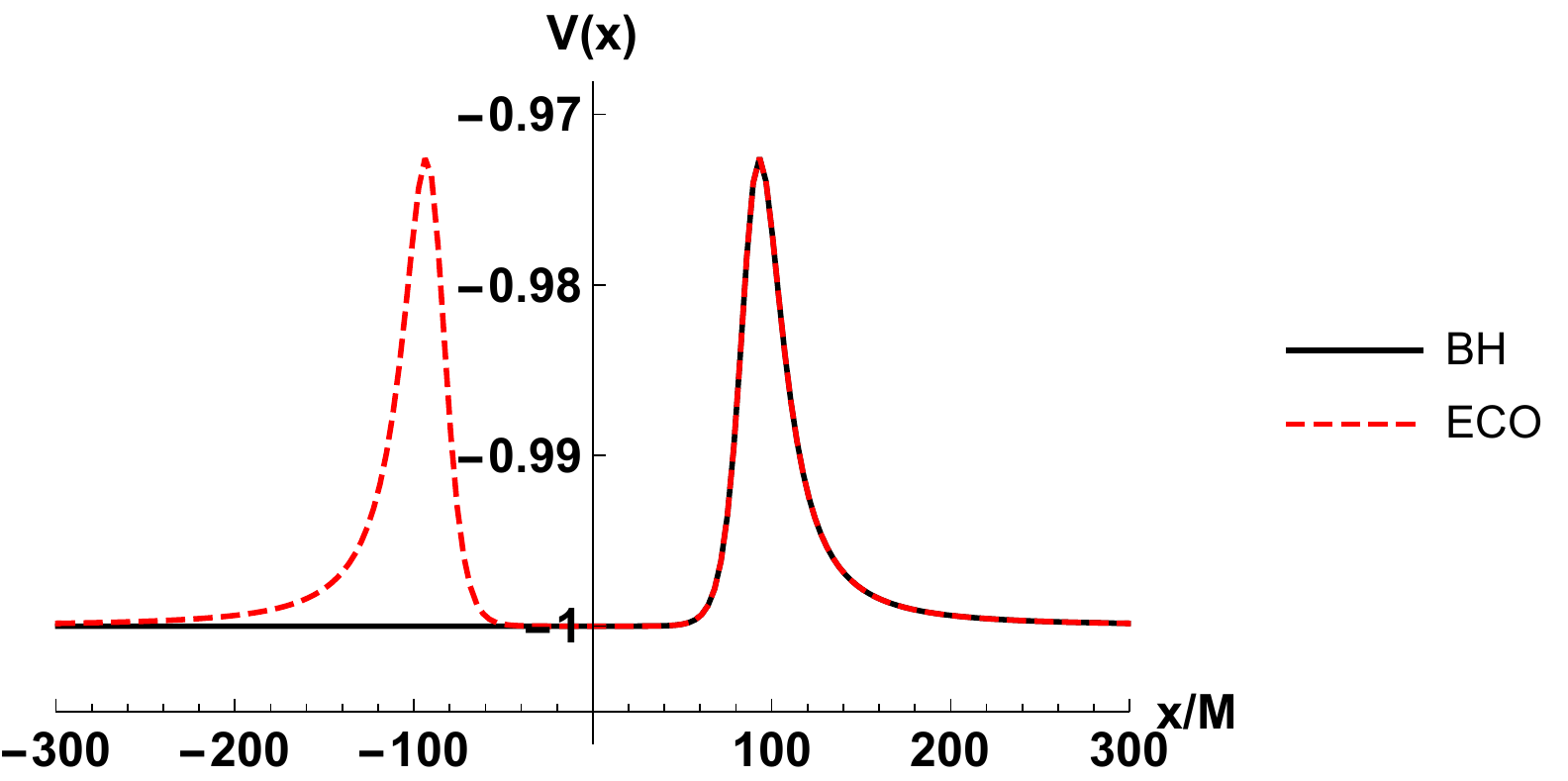}
 \caption[Effective potential]{The effective potential that a scalar wave sees in a (Schwarschild) black hole background (solid black line), and in a particular exotic compact, horizonless object or ECO (dotted red line). (In particular, this is the potential for a Solodukhin wormhole with $\lambda=10^{-10}$ \cite{Bueno:2017hyj,Dimitrov:2020txx}.) The coordinate $x$ is an appropriate tortoise coordinate such that $\lim_{r\rightarrow\infty} V(x(r)) = -1$. The potential that the scalar feels in the ECO background mimics that of the black hole background up until the horizon-scale structure at $x\sim -100M$. This structure leads to an effective cavity size of $L\sim  186M$.
 }
 \label{fig:echopotential}
\end{figure}

We can model the effects of the horizon-scale structure by a second ``bump'' in the potential that the wave perturbation feels --- see Fig. \ref{fig:echopotential}. In a black hole spacetime, there would be no extra bump and the perturbation is absorbed by the horizon, meaning any wave continues traveling towards the left on Fig. \ref{fig:echopotential}. In an ECO spacetime, the perturbation is trapped between the two potential bumps, and a small part of the wave ``leaks'' out every time the wave hits the rightmost bump --- this periodic leaking signal is precisely the echo. The time between echoes is twice the distance $L$ (in an appropriate tortoise coordinate) between these two bumps, and scales as $L\sim \log \epsilon$ when $\epsilon$ denotes the ``compactness'' of the ECO as in (\ref{eq:ECOepsilon}). In other words, $\epsilon$ can be \emph{exponentially} small (and the ECO correspondingly ultra-compact) without this echo time-scale becoming exponentially large! Even very compact microstructure can give measurable echoes in the ringdown phase.

Echoes of multi-centered bubbling microstate geometries were also discussed in \cite{Ikeda:2021uvc}. Note that it is not (yet) entirely clear how the behaviour of the echoes
will behave when the microstate geometry approaches the scaling limit. To have a \emph{clean} (i.e. distinguishable) echo, the potential well and bump structure in Fig. \ref{fig:echopotential} must also be relatively ``clean'' --- the bumps and wells must be clearly distinctive. A more realistic potential coming from a near-scaling microstate geometry will almost certainly be more chaotic, making the echo structure similarly chaotic and less distinguishable from background noise in the resulting observational signal. Further analysis of echoes in the scaling limit of microstate geometries is needed to understand this better.

\subsection{Black hole imaging}\label{sec:FBimaging}

\begin{figure}[ht]\centering
\includegraphics[width=\textwidth]{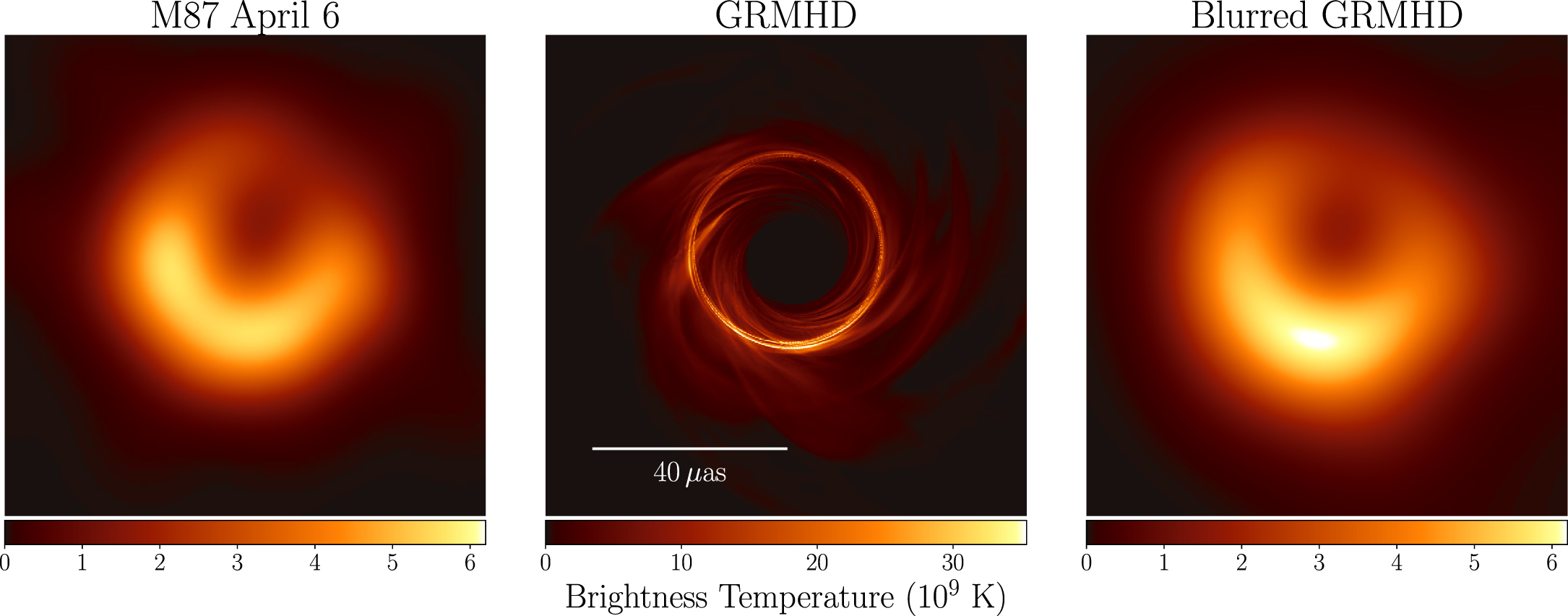}
\caption[EHT]{The black hole image as obtained by the EHT (left), together with a simulated image from a GRMHD simulation (middle), and the same simulated image after blurring to mimic the EHT telescope (right). The black hole shadow is the dark central region; the photon rings are clearly visible in the middle picture as the bright ring.\\ (This is Fig. 1 from \cite{EHT2019e}, licensed under \href{https://creativecommons.org/licenses/by/3.0/}{CC BY 3.0}.)}
\label{fig:EHT}
\end{figure}

Another recent observational advancement is \emph{black hole imaging} by the Event Horizon Telescope (EHT) --- an impressive collaboration of multiple telescopes throughout the world, creating a virtual Earth-sized telescope \cite{EHT2019a}. This allowes observations at a small enough wavelength to image the near-horizon region of the supermassive black hole at the center of the M87 galaxy (see Fig. \ref{fig:EHT}), and is expected to also be able to provide similar imaging of the black hole at the center of our galaxy.

Of course, a black hole --- being black --- does not emit any light itself; rather, the near-horizon environment of the black hole consists of a complicated plasma of accelerated, charged particles that emit light. This light is then affected by the black hole's geometry, bending around the black hole before escaping and finally arriving at the observing telescope. As such, this light probes and encodes the details of the near-horizon geometry of the black hole.

However, a big problem in disentangling the geometric details from the observed photons is the uncertainty in the physics of emission; many details of the plasma surrounding the black hole are still unknown. The images we currently have do not distinguish well between different models --- even very different ones \cite{EHT2019e}. This plasma uncertainty masks the details of the geometry, making it hard to extract precision geometrical data \cite{Gralla:2020pra}. Nevertheless, certain constraints on deviations from classical general relativity have already been calculated \cite{EventHorizonTelescope:2020qrl}. In the future, additional VLBI (Very Long Baseline Interferometry) such as improvements or additions to the current EHT telescopes, or space-based telescopes, would be able to observe the black holes at different wavelengths and ``see through'' the obscuring plasma effects, ameliorating this issue \cite{Haworth:2019urs,Gralla:2020srx}.

In a black hole image, there is always a black central region called the black hole \emph{shadow} (see Fig. \ref{fig:EHT}) --- this is due to the black hole horizon absorbing any light that travels too close to it. A horizonless object would \emph{not} absorb any light, so \emph{a priori} we may think that such an object must look transparent. This seems problematic if we also want a horizonless object to mimic a black hole, at least to some extent.

This issue was clarified for horizonless fuzzballs in \cite{Bacchini:2021fig}. There, images of multi-centered microstate geometries were discussed, and the mechanisms were elucidated how such a horizonless microstate geometry can still mimic the ``blackness'' of a black hole shadow. Light travelling close to the horizon scale will explore the microstructure sitting there. Such light will be trapped on \emph{very long lived, chaotic} orbits, so it takes a long time for it to escape the microstructure. In addition, as it explores the microstructure, it encounters \emph{regions of large tidal forces}.\footnote{See also the recent developments on \emph{stringy tidal forces} \cite{Martinec:2020cml,Bena:2020iyw}.} These large tidal forces imply that the light ray will cease to be well described by a non-interacting geodesic on this background; rather, the light ray will backreact on the geometry and perturb the microstate geometry. At the location of this perturbation, new photons can be emitted; however, since the regions of large tidal forces are also \emph{regions of large redshift}, the resulting photons that escape to the outside of the microstructure will necessarily be heavily redshifted to undetectable energies. In this way, the microstructure will effectively be rendered \emph{black}, because no un-redshifted photons will be able to escape from this region.

Finally, we can mention the possibility of imaging observables which can themselves ``bypass'' the obscuring effects of the photon emission details. \textbf{Photon rings} are an exciting candidate of such an observable \cite{Johnson:2019ljv}. The $n$-th photon ring is formed by photons which have travelled $n$ half-orbits around the black hole geometry before escaping the black hole environment. It can be shown that the photons in the $n$-th and the $(n+2)$-th photon rings must have been emitted from points that are extremely close to each other --- so even if the precise emission details are not known, the relative properties of these rings probe ``the same emission'' and so filter out the unknown details of this emission. It has been argued that the \emph{shape} of the $n=2$ photon ring can be very precisely measured with a proposed near-future space-based VLBI mission \cite{Gralla:2020srx} and that this would represent a precision test of the general relativistic black hole geometry. The \emph{relative intensities} of successive photon rings are also very sensitive probes to the geometry details, although it is unclear how well these would be measurable with (near) future experiments. In any case, understanding how microstate geometry photon rings can deviate from their black hole counterparts is still an exciting open question.

\section*{Acknowledgments}
I would like to thank the organizers of the 2021 Modave Summer School for inviting me to give these lectures, for giving me the chance to share this topic with the students present, and for being patient and flexible with the submission deadline for these notes.
I would also like to thank V. Dimitrov for taking nice notes during my lectures, especially including some interesting questions that were raised by the students.
I am supported by ERC Advanced Grant 787320 - QBH Structure, ERC Starting Grant 679278- Emergent-BH, and FWO Research Project G.0926.17N. This work is also partially supported by the KU Leuven C1 grant ZKD1118 C16/16/005.

\appendix
\addtocontents{toc}{\protect\setcounter{tocdepth}{1}}

\section{String Theory in One Minute}\label{sec:stringtheory}
This appendix is to give the unfamiliar reader a quick glance or overview --- a ``glossary'' as it were --- of some important string theory terms and concepts used throughout these notes.

There are many references on string theory and D-branes. Limiting myself to one freely available lecture note each, a nice introduction to string theory is David Tong's lecture notes \cite{Tong:2009np}; a comprehensive reference about D-branes is Clifford Johnson's book \cite{Johnson:2000ch}.

\paragraph{Strings; string and supergravity theories}
The fundamental object in string theory is the string --- also called an F1-string or F1-brane. There are closed strings (that form closed loops) and open strings.

There are two maximally symmetric string theories (i.e. with 32 supercharges): type IIA and type IIB. They are related by \emph{dualities}, which relate states in one of the theories to states in the other. The low-energy limit of type IIA/B string theory is type IIA/B supergravity in ten dimensions; this is typically the supergravity starting point for constructing solutions (such as microstate geometries). By performing \emph{dimensional reductions} over compact directions, we can obtain lower-dimensional supergravity theories from these ten-dimensional type IIA/B supergravity theories.

There is also an eleven-dimensional theory called M-theory. This can be seen as a kind of ``strong coupling'' limit of type IIA string theory. We only really have access to the low-energy limit of M-theory, which gives us the (unique) eleven-dimensional supergravity theory. It is also sometimes convenient to construct solutions in M-theory instead of type IIA/B (for example, see Section \ref{sec:braneinterpret}).

\paragraph{D-branes}
There are also many other extended objects in string theory called \textbf{D-branes}. Open strings must end on a D-brane. A D$p$-brane extends along $p$ spatial dimensions --- for example, a D1-brane is sometimes also called a ``D-string''. A D0-brane is then analogous to a point particle, having no spatial extent.

If we are are in type IIA string theory, then we have even-$p$ D-branes: D0, D2, D4, D6, D8. Type IIB has odd-$p$ D-branes: D1, D3, D5, D7, D9. (There are also D(-1)-branes; let's not go there.) The dualities between IIA and IIB theories map the different types of D$p$-branes into each other. M-theory has M2-branes and M5-branes --- when we dimensionally reduce M-theory to type IIA, these M-branes become D-branes and the F1-string. (Both type IIA and IIB string theories also have NS5-branes, which are the magnetic dual of F1-strings.)

\paragraph{Gauge field strengths}
A point particle is a source for a one-form potential with a two-form field strength --- for example, an electron in electromagnetism. In the same way, an object extending in $p$ spatial dimensions is a source for a $(p+1)$-form potential field with a $(p+2)$-form field strength. Concretely, D$p$-brane sources a $(p+2)$-form \emph{Ramond-Ramond} field strength $F_{(p+2)}$. An F1-string sources the Neveu-Schwarz three-form field strength $H_{(3)}$.

\paragraph{Kaluza-Klein compactifications}
To ``compactify'' or equivalently ``dimensionally reduce'' (or Kaluza-Klein reduce) along a dimension or direction means we essentially try to get rid of all dependence on this direction in our theory. For example, consider a metric in a $(d+1)$-dimensional theory:
\be \label{eq:compactmetric} ds_{(d+1)}^2 = X ds_d^2 + Y (dy+A)^2,\ee
where $y$ is a compact $S^1$ circular direction; $X,Y$ are scalar factors that do not depend on $y$, and $A$ is a one-form that also does not depend on $y$. We can then ``dimensionally reduce'' this \emph{family} of metrics to a $d$-dimensional theory, where the metric will be $ds_d^2$, and $A$ will give rise to an extra one-form gauge field in this lower-dimensional theory. (Note that a scalar also features in this story, and $X$ and $Y$ in (\ref{eq:compactmetric}) are related.)
A canonical reference on compactifications is \cite{popeKK}.

\paragraph{Supersymmetry and supersymmetric solutions}
Supersymmetry (or SUSY) is often colloquially described as the symmetry where each boson has a superpartner fermion (and vice versa). Concretely, a supersymmetric theory has an action which is invariant under an infinitesimal supersymmetry transformation, which is a transformation parametrized by a \emph{spinor}.

A supergravity theory is a gravity theory that is supersymmetric; the graviton (metric) has one or more superpartner(s), the gravitino field. Roughly, the number of superpartner gravitinos is the number of supersymmetries in the theory. There is a limit to the number of supersymmetries that a theory can have --- 32 --- and a theory that has this maximal amount is called \emph{maximally supersymmetric}. Type IIA/B and M-theory are all maximally supersymmetric supergravity theories.

We typically only consider bosonic solutions in supergravity, which means we turn all of the fermionic fields completely off (set them to zero). Such a solution in a supergravity theory is called a \emph{supersymmetric solution} if the solution is invariant under some supersymmetric transformations of the supergravity theory. Note the subtle difference between a supersymmetric theory and a supersymmetric solution: for the supergravity \emph{theory} to be supersymmetric, the theory must be invariant under a set of supersymmetric transformations for an \emph{arbitrary} spinor parameter; for a \emph{solution} in the theory to be supersymmetric, the solution must only be invariant under the same supersymmetry transformations for (at least) one particular spinor parameter. This particular spinor parameter is called a \emph{Killing spinor} of the solution. The amount of linearly independent Killing spinors that a given solution admits is the \emph{amount of supersymmetry} that a solution preserves.

Finding supersymmetric solutions is then equivalent to finding bosonic solutions for which the supersymmetry variations vanish. It can be shown that usually (i.e. when the so-called ``Bianchi identities'' are satisfied) a solution to the supersymmetry equations \emph{automatically} also solves the equations of motion. The supersymmetry equations are typically \emph{first-order} differential equations in the fields of the theory, so are typically easier to solve than the theory's (second-order) equations of motion.

\section{Quick Reference for Multi-centered Geometries}\label{sec:summarybubble}

\paragraph{General setup}
The five-dimensional supergravity action is:
\be S_5 = \int d^5x\, \left[\sqrt{-g}\left( R - \frac12 Q_{IJ} F^I\cdot F^J - Q_{IJ} \partial X^I\cdot \partial X^J\right) - \frac{1}{24}\epsilon^{\mu\nu\rho\sigma\lambda} A_\mu^I F_{\nu\rho}^{J} F_{\sigma\lambda}^{K}\right],\ee
with:
\begin{align}
 \frac16 C_{IJK} X^I X^J X^K &= 1,&
 Q_{IJ} &= \frac92 X_I X_J - \frac12 C_{IJK} X^K,&
 X_I &= \frac16 C_{IJK} X^J X^K.
\end{align}
We always take $C_{IJK} = |\epsilon_{IJK}|$ so that:
\begin{align} X^I &= \frac{Z}{Z_I}, & Z & = (Z_1Z_2Z_3)^{1/3}, & Q_{IJ} &= \frac12\delta_{IJ}(X^I)^{-2}.\end{align}
The metric and gauge fields are determined by:
\be \label{eqs:5Dmetric} ds_5^2 = -Z^{-2} (dt + k)^2 + Z ds_4^2, \qquad \Theta^I = dA^I + d\left( Z_I^{-1}(dt+k)\right).\ee

\paragraph{Gibbons-Hawking multi-centered solutions}
The four-dimensional base is of the Gibbons-Hawking form:
\be \label{eqs:GHbase} ds_4^2 = V^{-1}(d\psi + A)^2 + V ds_3^2, \qquad \vec{\nabla}\cross \vec{A} = \vec{\nabla} V. \ee
with $ds_3^2=dr^2+r^2d\theta^2+r^2\sin^2\theta d\phi^2$ as flat $\mathbb{R}^3$. Further, the two-forms are:
\be \Theta^I = \sum_{a=1}^3 \partial_a (V^{-1} K^I)\Omega_+^a,\ee
and all other functions are given by:
\begin{align}
\label{eqs:Z} Z_I &= \frac12 C_{IJK} \frac{K^J K^K}{V} + L_I,\\
 \label{eqs:mu} \mu &= \frac16 C_{IJK} \frac{K^I K^J K^K}{V^2} + \frac12 \frac{K^I L_I}{V} + M,\\
 \label{eqs:k} k &= \mu(d\psi + A)+\omega,\\
\label{eqs:omega} \vec{\nabla}\cross \vec{\omega} &= V \vec{\nabla} M - M \vec{\nabla} V + \frac12\left( K^I \vec{\nabla} L_I - L_I \vec{\nabla} K^I\right).
\end{align}
The solution is completely determined by the eight functions $H\equiv (V,K^I,L_I,M)$ which are \emph{harmonic on $\mathbb{R}^3$} ($\nabla^2 H=0$); these functions are then completely determined by their singularity structure or ``centers'':
\be H = h^0 + \sum_i \frac{h^i}{r_i},\ee
where $r_i = |\vec{r} - \vec{r}_i|$ is the (flat) $\mathbb{R}^3$ distance to the $i$-th center. The charge vector of a center is defined as:
\be \Gamma^i \equiv (v^i, k_I^i, l_I^i, m^i)=\left(v^i, k_1^i, k_2^i, k_3^i, l_1^i, l_2^i, l_3^i, m^i\right).\ee
The constant terms in the harmonic functions are the (asymptotic) moduli:
\be h^0 \equiv (v^0, k_I^0, l_I^0,m^0).\ee

\paragraph{Regularity}
Demanding no CTCs requires (to be satisfied everywhere):
\be \label{eqs:Q} \mathcal{Q} \equiv Z_1 Z_2 Z_3 V - \mu^2V^2 \geq 0,\qquad V Z_I \geq 0.\ee
Near a center, a necessary condition for no CTCs is to satisfy the bubble equations (there is one for each center $i$):
\be \label{eqs:bubble} \sum_{j\neq i} \frac{\langle \Gamma^i, \Gamma^j\rangle}{r_{ij}} = \langle h^0, \Gamma^i\rangle,\ee
where we have defined the intercenter distance $r_{ij} = |\vec{r}_i-\vec{r}_j|$ and the symplectic product:
\be \langle \Gamma^i, \Gamma^j\rangle \equiv \left(m^iv^j - \frac12 \sum_I k_I^i l_I^j\right) - (i\leftrightarrow j).\ee
If we want a \emph{smooth, horizonless} solution, then all centers must satisfy:
\be \label{eqs:lmsmooth} l_I^i = -\frac12 C_{IJK} \frac{k^i_Jk^i_K}{v^i}, \qquad m^i = \frac{1}{12} C_{IJK} \frac{k^i_Ik^i_Jk^i_K}{(v^i)^2}.\ee

\paragraph{Asymptotics and charges}
The moduli $h^0$ determine the asymptotics of the solution. The most typical asymptotic five-dimensional ($\mathbb{R}^{4,1}$) moduli are:
\be v^0 = k^0_I = 0, \qquad l^0_I = 1, \quad \sum_i v^i = 1,\ee
and $m^0$ is then determined by the sum of the bubble equations, $\langle h, \sum_i \Gamma^i\rangle = 0$. With these asymptotics, there are three electric charges and two angular momenta; for solutions that are \emph{completely smooth} these are given by:
\begin{align}
 Q_I &= -2 C_{IJK} \sum_i \frac{\tilde k^i_J \tilde k^i_K}{v^i}, & \tilde k^i_I &= k^i_I - v^i\sum_j k^j_I,\\
 J_R &= \frac43 C_{IJK} \sum_i \frac{\tilde k^i_I \tilde k^i_J \tilde k^i_K}{(v^i)^2},
\end{align}
and the expression for $J_L$ is more complicated expression (see eqs. (152)-(154) in \cite{Bena:2007kg}).

For four-dimensional asymptotics ($\mathbb{R}^{3,1}\times S^1$), we need the $\psi$ circle to become of constant size at infinity. One possible choice of moduli is:
\be v^0 = 1, \qquad k^0_I = 0, \qquad l^0_I = 1,\ee
and $m^0$ is determined by the sum of the bubble equations. Another possible choice is:
\be v^0 = l^0_I = 0, \qquad k^0_I = 1, \qquad m^0 = -\frac12,\ee
(as long as the sum of the bubble equations is satisfied). The general condition for having an asymptotically four-dimensional solution is $\lim_{r\rightarrow\infty}\mathcal{Q}=1$ (with $\mathcal{Q}$ defined in (\ref{eq:CTCQVZ})). Note that the four-dimensional metric is given by:
\be \label{eqs:4Dmetric} ds_\text{(4D)}^2 = -\mathcal{Q}^{-1/2} (dt+\omega)^2 + \mathcal{Q}^{1/2} ds_3^2.\ee

%
%
%
%
%
%
%
%
%
%

\section{Exercises and Solutions}\label{sec:exercises}
Below are a few exercises on multi-centered geometries, and one on superstrata. These exercises can be useful for a bubbling novice to get a good feeling for this geometries. Especially Exercise \ref{ex:scaling} on scaling geometries is important --- see e.g. Section \ref{sec:horizonscalebubbles}.

You may find assistance from Mathematica can be quite useful in some parts of these exercises.

The solutions are given below in Section \ref{sec:exsols}.

\medskip


\begin{exercise}[One-center black hole in 4D]\label{ex:singleBH}
Consider a solution with a single center at $r=0$ with charges:
\be \Gamma^1 = \left(0,q_1,q_2,q_3,0,0,0,q_0\right),\ee
with moduli:
\be h = \left( 0,1,1,1,0,0,0,-\frac12\right).\ee
\begin{exparts}
 \item Write down the 8 harmonic functions $H=(V,K^I,L_I,M)$ completely. Discuss $\omega, Z_I, \mu, k$. Is $\mathcal{Q}$ well-defined?
 \item Show that this is a black hole in four dimensions (with metric (\ref{eqs:4Dmetric})) by showing that $r=0$ (at a given time $t$) is a surface with a given finite area. (What is this area?) Are there conditions on $q_0,q_I$ to make this black hole physical?
\end{exparts}
\end{exercise}

\medskip

\begin{exercise}[Smooth centers]
Consider a smooth center at $r=0$ with arbitrary $v^1,k^1_I$ charges (and $l_I^1,m^1$ charges determined by the smoothness conditions). (There may also be other centers away from the origin in the system.)
\begin{exparts}
 \item Consider the $r\rightarrow 0$ behaviour of $Z_I$ and $\mu$ and show that the smoothness conditions indeed imply that these functions do not diverge at the center.
  \item Now, taking into account that there may be other centers away from the origin, show that satisfying the bubble equation for this center at $r=0$ is equivalent to demanding that $\mu\rightarrow 0$ at the center.
   \item Take $v^1=1$. Use the coordinate transformation $r=\rho^2/4$ to show that the metric at the center at $r=0$ looks simply like the origin of (flat) $\mathbb{R}^{4,1}$.
\end{exparts}
\end{exercise}

\medskip

\begin{exercise}[Two centers]
Consider two centers with arbitrary charges $\Gamma^1 = (v^1, k_I^1, l_I^1, m^1)$ and $\Gamma^2 = (v^2, k_I^2, l_I^2, m^2)$. The two centers are located on the $z$-axis at $z=\pm l/2$. Determine $l$ from the bubble equations and give any other conditions that the charges and moduli need to satisfy in order for the solution to be regular. Consider the case of arbitrary moduli $h = (v^0, k_I^0, l_I^0,m^0)$ as well as the specific moduli $h^{(1)} = (1,0,0,0,1,1,1,m_0)$ and $h^{(2)} = (0,1,1,1,0,0,0,-1/2)$.
\end{exercise}

\medskip

\begin{exercise}[Scaling solution] \label{ex:scaling}
Consider a three-center solution with harmonic functions:\footnote{This solution in inspired by the one used in \cite{Bacchini:2021fig}.}
\be \label{eq:3charmfuncs}
\begin{aligned}
    V &= \frac{1}{r_1} - \frac{1}{r_2}, & K^I &= 1 + P\left(\frac{1}{r_1} + \frac{1}{r_2}\right),\\
   L_I &= -P^2\left(\frac{1}{r_1} - \frac{1}{r_2}\right)  , &  M &= -\frac12 + \frac{P^3}{2}\left(\frac{1}{r_1} + \frac{1}{r_2}\right) + \frac{-q_0}{r_3},
\end{aligned}
\ee
where center $1,2$ are located at $x=y=0$ and $z=\pm l/2$, and center $3$ is located at $y=z=0$ and $x=R$. The bubble equations determine $l,R$ in terms of a single parameter $\lambda$ as:
\be \label{eq:lR3cent} l = 8 P^3 \lambda, \qquad  R = 2\lambda\sqrt{\frac{q_0^2}{(1-(1-3P^2)\lambda)^2}-4P^6}.\ee
This parameter $\lambda$ has an upper bound determined by the triangle inequalities\footnote{The bubble equations can be seen as determining the intercenter distances in terms of the center charges. These intercenter distances together must satisfy the triangle inequalities in order for the three centers to lie in a triangle, since otherwise there can be no solution with the given intercenter distances.} (or equivalently here, demanding that $R>0$). The lower bound is simply zero. We will consider the $\lambda\rightarrow 0$ limit, called the \emph{scaling limit} of the solution. In this limit, all three centers converge on the origin, $r_i\rightarrow r$.
\begin{exparts}
 \item Confirm that centers $1,2$ are smooth centers.\footnote{Center 3 is clearly \emph{not} a smooth center, only having a charge in the $M$ channel. In the 10D IIA frame where the smooth centers are fluxed D6 branes, center 3 is a (stack of) D0 brane(s).}
 \item Expand the harmonic functions to zeroth order in $\lambda$. Compare to the harmonic functions of the single center black hole of Exercise \ref{ex:singleBH} to conclude that the scaling limit approaches the black hole geometry.
 \item  Show that the proper distance between centers 1 and 2 tends to a \emph{finite} value as $\lambda\rightarrow 0$.
 \item Solve the bubble equations yourself with the charges given in (\ref{eq:3charmfuncs}) to find the intercenter distances $l_{12},l_{13},l_{23}$ in terms of the charges and to confirm the expressions for $l,R$ given above.
\end{exparts}
\end{exercise}

\medskip

\begin{exercise}[Black hole deconstruction]
  Consider the setup from the previous exercise, but instead of the third center, consider an arbitrary amount of centers of the same species as the third center (i.e. only having an $M$ charge), so that the total ($M$) charge of all these new centers together is $q_0$. Show, using the new bubble equations, that one possible configuration for these new centers is that they all sit at arbitrary positions on a ring in the $z=0$ plane with radius $R$ given by (\ref{eq:lR3cent}).
\end{exercise}

\medskip

\begin{exercise}[Three-dimensional metric limits of superstrata]
Take a general multi-mode $(1,0,n)$ superstrata geometry. We can rewrite the metric as:
\be ds_6^2 = d\hat{s}^2_\mathcal{K} + d\hat{s}^2_{S^3},\ee
where:
\begin{align}
 d\hat{s}^2_{S^3} &= g_{\theta\theta}d\theta^2 + g_{11}(d\varphi_1 + A_t^{(1)}dt)^2 + g_{22}(d\varphi_2 + A_t^{(2)}dt + A_y^{(2)}dy)^2,\\
 d\hat{s}^2_\mathcal{K} &= \hat{g}_{tt} dt^2 + \hat{g}_{yy}(dy + \hat{A}_t^{(y)}dt)^2 + \hat{g}_{rr}dr^2.
 \end{align}
 We will find $d\hat{s}^2_\mathcal{K}$ and study some of its limits.
\begin{exparts}
 \item Find all of the metric components $g_{\theta\theta}, g_{11},g_{22}$ and the off-diagonal components $A_t^{(1)},A_t^{(2)},A_y^{(2)}$. Then find $\hat{g}_{tt},\hat{g}_{yy}, \hat{g}_{rr}$ and $\hat{A}_t^{(y)}$.
 \item Now that you found $d\hat{s}^2_\mathcal{K}$, consider it at infinity, $r\rightarrow \infty$. Show that, after a radial coordinate redefinition and a gauge transformation, it reduces to $AdS_3$ in its canonical form at leading order:
 \be ds_{AdS_3}^2 = R^2 \left[ \frac{d\rho^2}{\rho^2} + \rho^2 dy'^2  - \rho^2 dt'^2\right] + O(\rho^0).\ee
 \item Now let's consider the metric $d\hat{s}^2_\mathcal{K}$ near the ``cap'' $r=0$. It is convenient to use the radial coordinate redefinition $\rho=r/a$. Expand the metric to $\mathcal{O}(\rho^2)$ to conclude that there is no conical singularity and that the metric limits precisely to flat Minkowski space at $r=\rho=0$.
\end{exparts}
\end{exercise}


\subsection{Exercise solutions}\label{sec:exsols}
I will choose to refer to the equations in the quick reference of appendix \ref{sec:summarybubble}.

\medskip

\begin{exsol}[One-center black hole in 4D]
\begin{solparts}
 \item 
Clearly, $V=L_I=0$. Further:
\be \label{eq:1cBHKM} K^I = 1 + \frac{q_I}{r}, \qquad M = -\frac12 +\frac{q_0}{r}.\ee
From (\ref{eqs:omega}), it is clear that $\omega=0$. (Remember, the integration constant from integrating (\ref{eqs:omega}) is unimportant and can be absorbed in a redefinition of the coordinate $t$.) It is also clear from (\ref{eqs:Z}), (\ref{eqs:mu}), (\ref{eqs:k}) that the $Z_I,\mu,k$ quantities are ill-defined since $V=0$. However, $\mathcal{Q}$ of (\ref{eqs:Q}) is finite. The easiest way to see this is to take $V=\epsilon$, compute $\mathcal{Q}$ in (\ref{eqs:Q}), and then take $\epsilon\rightarrow 0$. This gives us:
\be \mathcal{Q} = Z_1Z_2Z_3 V - \mu^2 V^2 = -2\left( 1 + \frac{q_1}{r}\right)\left( 1 + \frac{q_2}{r}\right)\left( 1 + \frac{q_3}{r}\right)\left( -\frac12 + \frac{q_0}{r}\right).\ee
\item At $r\sim 0$ and at constant $t$, we have:
\be ds^2 \sim \left(\sqrt{2}\sqrt{-q_0q_1q_2q_3} \right) (d\theta^2 + \sin^2\theta d\phi^2).\ee
So at $r=0$, the two-sphere parametrized by $(\theta,\phi)$ --- the horizon of the black hole --- has a finite area. The radius squared of this two-sphere is $\sqrt{2}\sqrt{-q_0q_1q_2q_3} $, so its area is simply $A_\text{BH} = 4\sqrt{2}\pi \sqrt{-q_0q_1q_2q_3}$. This area should be positive, so we must have:
\be q_0 q_1 q_2 q_3 <0.\ee
Typically, one takes $q_0<0$ and $q_{1,2,3}>0$.
\end{solparts}
\end{exsol}

\medskip

\begin{exsol}[Smooth centers]
 Clearly, we have:
 \be \label{eq:ex2H} H = h^0 + \frac{\Gamma^1}{r},\ee
 where $h^0$ is not specified, and the only relation that the $\Gamma^1=(v^1,k_I^1,l_I^1,m^1)$ must satisfy is (\ref{eqs:lmsmooth}), so:
 \be \label{eq:ex2smooth} l_1^1 = - \frac{k_2^1 k_3^1}{v^1}, \quad l_2^1 = - \frac{k_1^1 k_3^1}{v^1}, \quad l_3^1 = - \frac{k_1^1 k_2^1}{v^1}, \quad m^1 = \frac12 \frac{k_1^1k_2^1k_3^1}{(v^1)^2}.\ee
 \begin{exparts}
  \item At $r\rightarrow 0$, we can ignore the moduli $h^0$ in (\ref{eq:ex2H}). The first term in the expression (\ref{eqs:Z}) for $Z_1$ will behave as:
  \be \frac{K^2 K^3}{V} = \frac{k^1_2k^1_3}{r^2} \left( \frac{v^1}{r}\right)^{-1} + \mathcal{O}(r^0) = \frac{k^1_2k^2_3}{v^1} r^{-1} + \mathcal{O}(r^0),\ee
  and the second term simply as:
  \be L_1 = l^1_1 r^{-1} + \mathcal{O}(r^0).\ee
  The diverging $\sim r^{-1}$ part of $Z_1$ will then vanish if and only if:
  \be\label{eq:prevsmooth} l^1_1 +  \frac{k^1_2k^2_3}{v^1} = 0,\ee
  which of course is precisely the smoothness condition (\ref{eq:ex2smooth}). The analysis for $Z_2,Z_3$ is analogous. For $\mu$, staring at (\ref{eqs:mu}) should hopefully also make it clear that the only possible divergence goes as $\sim r^{-1}$. Now, there are three such diverging terms; however, the $K^IL_I/V$ term combines with the $K^IK^JK^K/V^2$ term due to the previous smoothness condition (\ref{eq:prevsmooth}). When the dust settles, the condition for the $\mathcal{O}(r^{-1})$ term in $\mu$ to vanish is simply (\ref{eq:ex2smooth}).
  
  \item As we derived above, the smoothness conditions (\ref{eq:ex2smooth}) tell us that $\mu \sim \mathcal{O}(r^0)$ as $r\rightarrow 0$. Let's now calculate this $\mathcal{O}(r^0)$ piece. First, we note that we can expand all the harmonic functions as:
  \be \label{eq:Hexp} H = \frac{\Gamma^1}{r} + \left( h^0 + \sum_{i > 1} \frac{\Gamma^i}{r_{1i}} \right) + \mathcal{O}(r),\ee
  where the sum is over all other centers, and $r_{1i}$ is the distance between the first center (at $r=0$) and the $i$-th center. The expansion (\ref{eq:Hexp}) allows us to pick out the $\mathcal{O}(r^0)$ terms in (\ref{eqs:mu}). Note first that:
  \be V^{-2} = \frac{r^2}{(v^1)^2} - \frac{2}{(v^1)^3} \left( v^0 + \sum_{i > 1} \frac{v^i}{r_{1i}} \right) r^3 + \mathcal{O}(r^4).\ee
  Then, we have:
  \be \frac{K^1 K^2 K^3}{V^2} = -2\frac{k_1^1 k_2^1 k_3^1}{(v^1)^3}\left( v^0 + \sum_{i > 1} \frac{v^i}{r_{1i}} \right) + \left[ \frac{k_2^1k_3^1}{(v^1)^2}\left( k^0_1 +  \sum_{i > 1} \frac{k_1^i}{r_{1i}} \right) + (\text{cyclic in }K^1,K^2,K^3)\right].\ee
  We can use the smoothness conditions (\ref{eq:ex2smooth}) to rewrite this as:
  \be \frac{K^1 K^2 K^3}{V^2} = -4 \frac{m^1}{v^1} \left( v^0 + \sum_{i > 1} \frac{v^i}{r_{1i}} \right) - \left(\sum_I\right)\frac{l_I^1}{v^1}\left( k^0_I +  \sum_{i > 1} \frac{k_I^i}{r_{1i}} \right).\ee
  Similarly, we can find that:
  \be \frac12 \frac{L_I K^I}{V} = + 3 \frac{m^1}{v^1}\left( v^0 + \sum_{i > 1} \frac{v^i}{r_{1i}} \right) + \frac12 \left(\sum_I\right)\frac{k_I^1}{v^1}\left( l^0_I +  \sum_{i > 1} \frac{l_I^i}{r_{1i}} \right) + \frac12 \left(\sum_I\right)\frac{l_I^1}{v^1}\left( k^0_I +  \sum_{i > 1} \frac{k_I^i}{r_{1i}} \right).\ee
  Combining these expressions with $M$, and multiplying by $v^1$, we find:
  \begin{align}
  v^1 \mu &= -m^1  \left( v^0 + \sum_{i > 1} \frac{v^i}{r_{1i}} \right) + v^1 \left( m^0 + \sum_{i > 1} \frac{m^i}{r_{1i}} \right)\\
   &  + \frac12 \left(\sum_I\right) \left[ k^1_I \left( l^0_I +  \sum_{i > 1} \frac{l_I^i}{r_{1i}} \right) - l^1_I \left( k^0_I +  \sum_{i > 1} \frac{k_I^i}{r_{1i}} \right) \right] + \mathcal{O}(r).
   \end{align}
   Demanding that $\mu$ vanishes as $r\rightarrow 0$ is then equivalent to demanding:
   \be \sum_{i>1} \frac{ m^1 v^i-v^1 m^i  + \frac12 \sum_I( l^1_I k^i_I-k^1_I l^i_I)}{r_{1i}} =m^0 v^1- v^0 m^1   + \frac12 \sum_I(l_I^0k_I^1-k_I^0 l^1_I),\ee
   which is precisely the bubble equation (\ref{eqs:bubble}) for center $1$:
   \be \sum_{i>1} \frac{\langle \Gamma^1,\Gamma^i\rangle}{r_{1i}} = \langle h^0, \Gamma^1\rangle.\ee
  
  \item (This is described in Section 4.1 of \cite{Bena:2007kg}.)
  From the above considerations, we now that as $r\rightarrow 0$, the $Z_I$'s go to a constant, $Z_I\rightarrow Z_I^c$, and $\mu\rightarrow 0$. Further, although it is not immediately obvious from (\ref{eqs:omega}), the bubble equations also ensure that $\omega\sim 0$ as $r\rightarrow 0$ (proving this is a good extra exercise!). Then, for the Gibbons-Hawking base (\ref{eqs:GHbase}), when $V\sim 1/r$ (near the center), we have $A\sim \cos\theta d\phi$. So, this means the metric (\ref{eqs:5Dmetric}) will look like:
  \be ds_5^2 = -Z_c^{-2} dt^2 + Z_c \left( r (d\psi +\cos\theta d\phi)^2 + r^{-1}\left[ dr^2+ r^2 d\theta^2 + r^2\sin^2\theta d\phi^2\right]\right) .\ee
  Finally, the suggested coordinate transformation $r= \rho^2/4$ gives us:
  \be ds_5^2 = -Z_c^{-2} dt^2 + Z_c  \left(  d\rho^2 +    \frac{\rho^2}{4}\left[   d\theta^2 + \sin^2\theta d\phi^2+(d\psi +\cos\theta d\phi)^2\right] \right).\ee
  The constant $Z_c$ can be absorbed in a redefinition of the time coordinate $t$ to give an overall factor for the entire metric. This metric is now simply the metric of \emph{five-dimensional flat space} (the angular part is precisely that of an $S^3$ with radius $\rho$, using Hopf fibration coordinates). Near $\rho\sim 0$ (i.e. $r\sim 0$), the metric is singular, but this is a simple coordinate singularity due to $\rho=0$ being an origin of $\mathbb{R}^{4,1}$. A coordinate singularity due to an origin is of course not a physical singularity, so we can conclude that $r\rightarrow 0$ is a \emph{smooth, non-singular point} of the geometry.
 \end{exparts}
\end{exsol}

\medskip

\begin{exsol}[Two centers]
The bubble equations (\ref{eqs:bubble}) for two centers can be written together as:
 \be \frac{\langle \Gamma^1, \Gamma^2\rangle}{r_{12}} = \langle h^0, \Gamma^1\rangle = -\langle h^0, \Gamma^2\rangle .\ee
 The last inequality is from the second bubble equation. The moduli must then satisfy:
 \be \label{eq:exh0}\langle h^0, \Gamma^1 + \Gamma^2\rangle = 0.\ee
 The intercenter distance $r_{12} = l$ is determined by the bubble equation(s) as:
 \be l = \frac{\langle \Gamma^1, \Gamma^2\rangle}{\langle h^0, \Gamma^1\rangle}.\ee
 Finally, both centers must satisfy the smoothness conditions (\ref{eqs:lmsmooth}) if the solution is to be smooth. 
 For the specific moduli $h^{(1)} = (1,0,0,0,1,1,1,m_0)$, we can use (\ref{eq:exh0}) to determine $m_0$:
 \be m_0 =\frac{1}{\sum_{i=1,2}v^i}\left( \sum_{i=1,2} m^i-\frac12 \sum_I \sum_{i=1,2} k_I^i\right).\ee
 For the moduli $h^{(2)} = (0,1,1,1,0,0,0,-1/2)$, we must instead restrict the charges due to (\ref{eq:exh0}):
 \be\label{eqs:sumbubble} \sum_{i=1,2}\left( v^i +\sum_I l_I^i\right) = 0.\ee
\end{exsol}

\medskip

\begin{exsol}[Scaling solution]
I will choose to work with the five-dimensional metric (\ref{eqs:5Dmetric}); an alternative would be to work with the four-dimensional metric (\ref{eqs:4Dmetric}), since these solutions are asymptotically four-dimensional ($\mathbb{R}^{3,1}\times S^1$).
  \begin{exparts}
   \item This is a simple matter of confirming (\ref{eqs:lmsmooth}) for the center charge vectors:
   \be \Gamma^1 = \left( 1, P, P, P, -P^2, -P^2, -P^2, \frac{P^3}{2}\right), \qquad \Gamma^2 = \left( -1, P, P, P, P^2, P^2, P^2, \frac{P^3}{2}\right).\ee
   
   \item Anything proportional to $r_1^{-1}-r_2^{-1}$ will tend to zero (to $\mathcal{O}(\lambda^0)$) while $r_1^{-1}+r_2^{-1}\rightarrow 2r^{-1}$. This means that, to $\mathcal{O}(\lambda^0)$, the harmonic functions are precisely given by the same form as the single-center black hole ones (\ref{eq:1cBHKM}), with $(q_0)_\text{(BH)}= P^3-q_0$ and $(q_I)_\text{(BH)} = 2P$.

   \item (This reasoning is also sketched in Section 8.5 of \cite{Bena:2007kg}.) The \emph{proper distance} $d_{12}$ between the two centers is simply the integral:
   \be d_{12} =  \int_{z=-l/2}^{z=+l/2}dz\,   \sqrt{g_{zz}} ,\ee
   which we evaluate at $x=y=0$ (and $\psi$ constant).
   Now, $g_{zz} = -Z^{-2}k_z^2 + Z V$. We note that the $z$ direction is perpendicular to $d\psi$; moreover, for our configuration $\omega\sim d\phi$ and $A\sim d\phi$ (this is a good extra exercise to show this!) so that $\omega_z=A_z=0$. Then, remembering that $k = \mu(d\psi + A) + \omega$, we conclude that $k_z=0$. So we have:
\be
 d_{12} =  \int_{-l/2}^{+l/2}dz\,  (Z_1Z_2Z_3 V^3)^{1/6} = \int_{-l/2}^{+l/2}dz\,  \sqrt{\frac{(l+4 P-2 z) (l+4 P+ 2z)}{l^2-4z^2}}.
\ee
Now, in the scaling limit, $l\sim \lambda$ and $|z|\leq l/2$ as $\lambda\rightarrow 0$. This means that the numerator will be dominated by $(4P)^2$, so:
\be d_{12}|_\text{(scaling)} =\lim_{l\rightarrow 0} \int_{-l/2}^{+l/2}dz\, 4P  \sqrt{\frac{1}{l^2-4z^2}} = \lim_{l\rightarrow 0} 2  \pi P = 2\pi P.\ee
So in the scaling limit, the proper distance between centers $1$ and $2$ tends to $2\pi P$. 

\item The bubble equations are:
\be \frac{\langle \Gamma^1,\Gamma^2\rangle}{r_{12}} + \frac{\langle \Gamma^1,\Gamma^3\rangle}{r_{13}} = \langle h^0,\Gamma^1\rangle,\qquad \frac{\langle \Gamma^1,\Gamma^3\rangle}{r_{13}} + \frac{\langle \Gamma^2,\Gamma^3\rangle}{r_{23}} = -\langle h^0,\Gamma^3\rangle.\ee
There is also a third bubble equation (for center 2), but since the sum of the bubble equations is satisfied (check that (\ref{eqs:sumbubble}) is satisfied!), this third bubble equation will automatically be satisfied if these two bubble equations are. These bubble equations simplify to:
\be \label{eq:bubblesscaling} \frac{4P^3}{l} - \frac{q_0}{r_{13}} = \frac12 - \frac32P^2, \qquad \frac{q_0}{r_{13}} - \frac{q_0}{r_{23}} = 0.\ee
The second equation tells us that the triangle is isosceles, $r_{13}=r_{23}$. Then, defining $\lambda$ by setting $l = 8P^3\lambda$, we find:
\be \label{eq:r13} r_{13} = r_{23} = \frac{2q_0 \lambda}{1- (1-3P^2)\lambda}.\ee
Since we put centers $1$ and $2$ at $z=\pm l/2$, this implies the third center must be located on the equatorial plane $z=0$. Further, (\ref{eq:r13}) implies that it has to lie on the circle of radius $R^2 = r_{13}^2-(l/2)^2$; this gives precisely the $R$ value in (\ref{eq:lR3cent}).

There are no further equations to solve,\footnote{Although note that we require $q_0>P^3$ so that $(q_0)_\text{(BH)}<0$, see above and Exercise \ref{ex:singleBH}.} so $\lambda$ is a \emph{free parameter} of the solution. However, not any value of $\lambda$ is allowed. First of all, intercenter distances should be positive, so $\lambda>0$ (assuming that $P>0$). Further, the intercenter distances $r_{13}=r_{23}$ and $r_{12}=l$ only can define a triangle if the \emph{triangle inequalities} are satisfied:
\be\label{eqs:triangle} r_{12} + r_{13} > r_{23}, \qquad |r_{12}-r_{13}| < r_{23}, \qquad \text{(and cyclic in all sides)}.\ee
(If the triangle inequalities are not satisfied for these distances, it is impossible to construct a triangle in Cartesian space with the three lengths given.)
The only non-trivial triangle inequalities for this triangle are the two given in (\ref{eqs:triangle}). It is not obvious to solve these in full generality. An example region of allowed solutions for fixed $P=2$ is:
\be P = 2, \qquad q_0 > 16 \quad \text{and} \quad 0<\lambda < \frac{1}{176}(q_0-16).\ee
In principle, one should also check for a given solution that (\ref{eqs:Q}) is satisfied so that there are no CTCs.
  \end{exparts}
\end{exsol}

\medskip

\begin{exsol}[Black hole deconstruction]
 Instead of only one center with $\Gamma^3=(0^7,-q_0)$, let's consider a number of centers with charge vectors $\Gamma^\alpha = (0^7,-q_\alpha)$ with $\sum_\alpha q_\alpha = q_0$. In this case, the sum of the bubble equations is still vanishing, and the other bubble equations are an altered version of (\ref{eq:bubblesscaling}):
 \be \frac{4P^3}{l} - \sum_\alpha \frac{q_\alpha}{r_{1\alpha}} = \frac12 - \frac32P^2, \qquad \frac{q_\alpha}{r_{1\alpha}} - \frac{q_\alpha}{r_{2\alpha}} = 0,\ee
 where the second equation is valid for each $\alpha$. The reason that the bubble equations are so ``easily'' modified is essentially because $\langle \Gamma^\alpha, \Gamma^{\alpha'}\rangle = 0$ for any $\alpha,\alpha'$ --- all of these extra centers don't ``talk to each other'' in the bubble equations. Once again, we see that all of the $\alpha$ centers must lie in the equatorial plane as $r_{1\alpha} = r_{2\alpha}$. The first bubble equation tells us that:
 \be \sum_\alpha \frac{q_\alpha}{r_{1\alpha}} = \frac{q_0}{r_{13}},\ee
 with $r_{13}$ as given in the previous exercise. Since $\sum_\alpha q_\alpha = q_0$, a simple solution to this equation is to put $r_{1\alpha} = r_{13}$ for all $\alpha$. This means all of the $\alpha$ centers must sit on a ring in the equatorial plane ($z=0$) with radius $x^2+y^2 = R^2 = r_{13}^2-(l/2)^2$. There is no further restriction on their placement; all of the $\alpha$ centers can be placed on arbitrary places on this ring.
 
 This problem is inspired by the black hole deconstruction paradigm \cite{Denef:2007yt}.
\end{exsol}

\medskip

\begin{exsol}[Three-dimensional metric limits of superstrata]
 This is described in Sections 4.1 and 4.2 of \cite{Heidmann:2019xrd}; see especially eqs. (4.3), (4.4), (4.6), (4.8) therein.
\end{exsol}

\bibliographystyle{toine}
\bibliography{fuzzballobservations}

\end{document}